\def\kms{km s$^{-1}$}
\def\hi{H{\sc i}}
\def\hii{H{\sc ii}}
\def\msun{M$_\odot$}

\def\mjyb{mJy beam$^{-1}$}

\def\radec{RA, Dec.(J2000)}

\def\coc{$^{13}$CO(J=2$\rightarrow$1)}

\documentclass[printer]{aa}  

\usepackage{natbib}
\bibpunct{(}{)}{;}{a}{}{,}
\usepackage{graphicx}
\usepackage{amssymb,amsmath}
\usepackage{txfonts}
\begin{document}
\title{Star forming regions linked to RCW\,78 and the discovery of a new IR bubble}

\author{C.E. Cappa\inst{1,2}\fnmsep\thanks{Member of 
          Carrera del Investigador, CONICET, Argentina}
          \and M. Rubio\inst{3}
          \and G.A. Romero\inst{1}\fnmsep\thanks{Member of 
          Carrera del Investigador, CONICET, Argentina}
          \and N.U. Duronea\inst{2}
          \and V. Firpo\inst{1,4}
                    }

\institute{Facultad de Ciencias Astron\'omicas y Geof\'{\i}sicas, Universidad Nacional de La Plata, Argentina \\ \email{ccappa@fcaglp.unlp.edu.ar}
          \and Instituto Argentino de Radioastronom\'{\i}a, CCT-La Plata, CONICET, C.C.5., 1894, Villa Elisa, Argentina
          \and Departamento de Astronom\'{\i}a, Universidad de Chile, Casilla 36-D, Santiago, Chile   
          \and Instituto de Astrof\'{\i}sica de La Plata, CCT-La Plata, CONICET, Paseo del Bosque s/n, La Plata, Argentina
             }

   \date{Received September 15, 1996; accepted March 16, 1997}

 
  \abstract
   {}
   {With the aim of investigating the presence of molecular and dust clumps linked to two star forming regions identified  in the expanding molecular envelope of the stellar wind bubble RCW\,78, we analyzed the distribution of the molecular gas and cold dust.  }
   {To accomplish this study we performed dust continuum observations at 870 $\mu$m  and $^{13}$CO(2-1) line observations with the APEX telescope, using LABOCA and SHeFI-1 instruments, respectively, and analyzed Herschel images at 70, 160, 250, 350, and 500 $\mu$m.}  
   {These observations allowed us to identify cold dust clumps linked to region B (named the Southern clump) and region C (clumps 1 and 2) and an elongated Filament. Molecular gas was clearly detected linked to the Southern clump and the Filament. The velocity of the molecular gas
is compatible with the location of the dense gas in the expanding envelope of RCW\,78. We estimate  dust temperatures and total masses for the dust condensations from  the emissions at different wavelengths in the far-IR and from the molecular line using LTE and the virial theorem. Masses obtained through different methods agree within a factor of 2-6. CC-diagrams and SED analysis of young stellar objects confirmed the presence of intermediate and low mass YSOs in the dust regions, indicating that moderate star formation is present.  In particular, a cluster of IR sources was identified inside the Southern clump.

The IRAC image at 8 $\mu$m revealed the existence of an infrared dust bubble of 16\arcsec\ in radius probably linked to the O-type star HD\,117797 located at 4 kpc. The distribution of the near and mid infrared emission  indicate that warm dust is  associated with the bubble.    
}
   {}
  
   \maketitle
%
\section{Introduction}

It is well established that objects at the first stages of star formation are inmersed in dust and cold  dense gas from their natal cloud. Dense molecular material can be found in the expanding envelopes surrounding \hii\ regions and stellar wind bubbles (SWB) (\citealt{zavagnoetal05}, and references therein). Thus, the dense gas shells that encircle these structures are  potential sites for the formation of new stars. In fact, many studies have shown the presence of active areas of star formation in the environs of these structures \citep[e.g.][]{deharvengetal10,zhang+wang12}.

Two physical processes have been proposed for the onset of star formation in the outer dense shells of these expanding structures: the  {\it collect and collapse} and the {\it radiatively driven implossion} (RDI) mechanisms. In the {\it collect and collapse} model the dense shell originated in the expansion of the ionized region becomes unstable and fragments, leading to the formation of massive stars \citep{elmegreen+lada77}, while the RDI process involves the compression of pre-existent overdensities in the molecular gas due to the expansion of the ionized region, leading to the formation of low and intermediate mass stars \citep{sandfordetal82,lefloch+lazareff94}.

Observational evidence of these star forming processes includes the existence of dense and neutral gas layers surrounding the ionized regions and the presence of high density clumps. Protostars are enshrouded in dense and cold  molecular clumps and dust cocoons present in the envelopes, in regions characterized by high extinction \citep{deharvengetal08}. Optically thin sub-millimeter  continuum emission from dust allows dust emission peaks associated with the dense molecular clumps to be found. In this context,  kinematical information of the molecular gas linked to the dust clumps can help  to confirm the association of the dust cocoons with  the dense molecular layer surrounding SWBs. $^{13}$CO is an optically thin tracer that can reveal areas  of high molecular gas column density and allows to study the kinematics of the  molecular clumps, although it may  fail to probe the densest molecular cores because it freezes onto dust grains at high densities \citep{massietal07}.

\begin{figure}
\centering
\includegraphics[width=8.5cm]{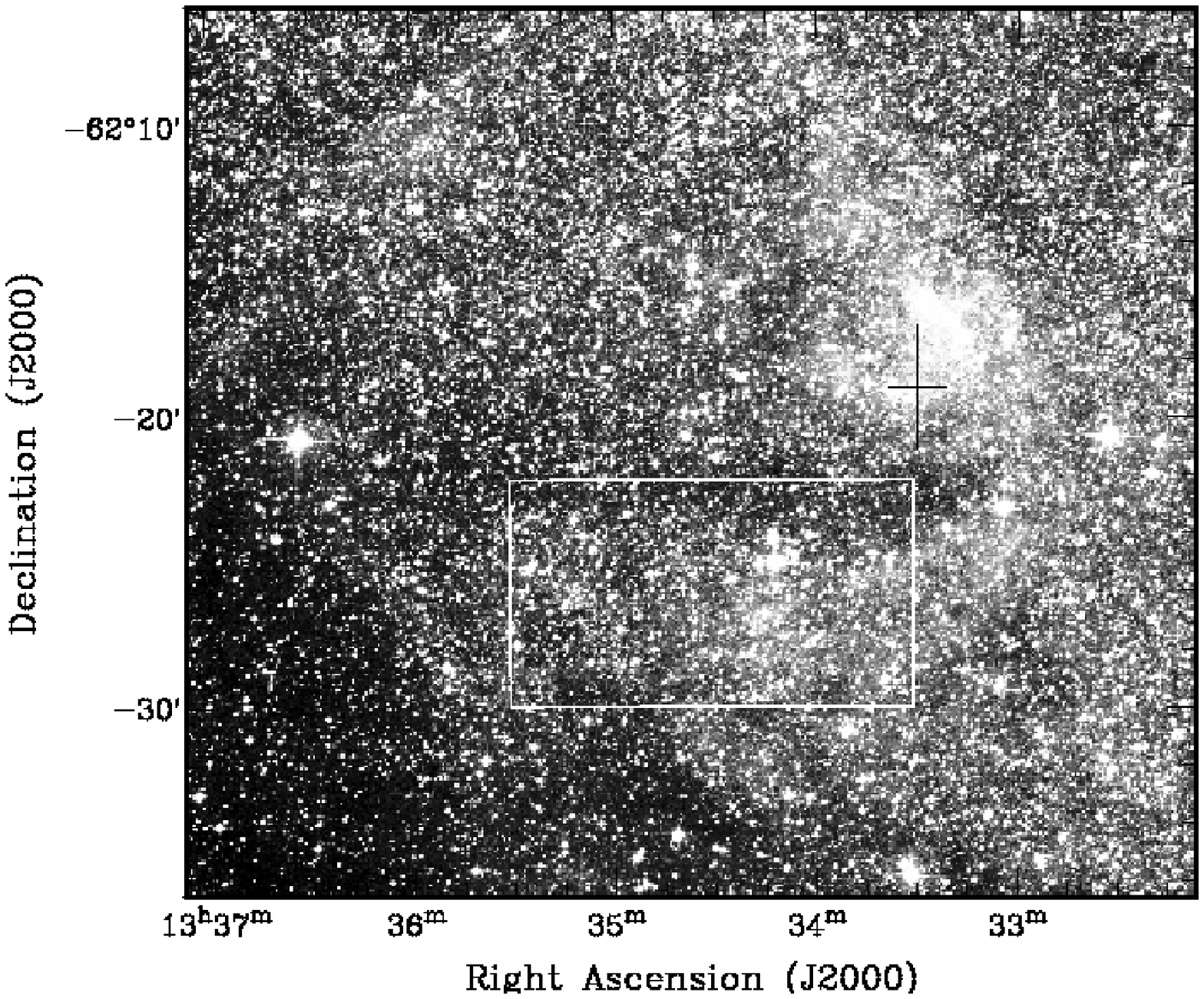}
\includegraphics[width=6cm,angle=-90]{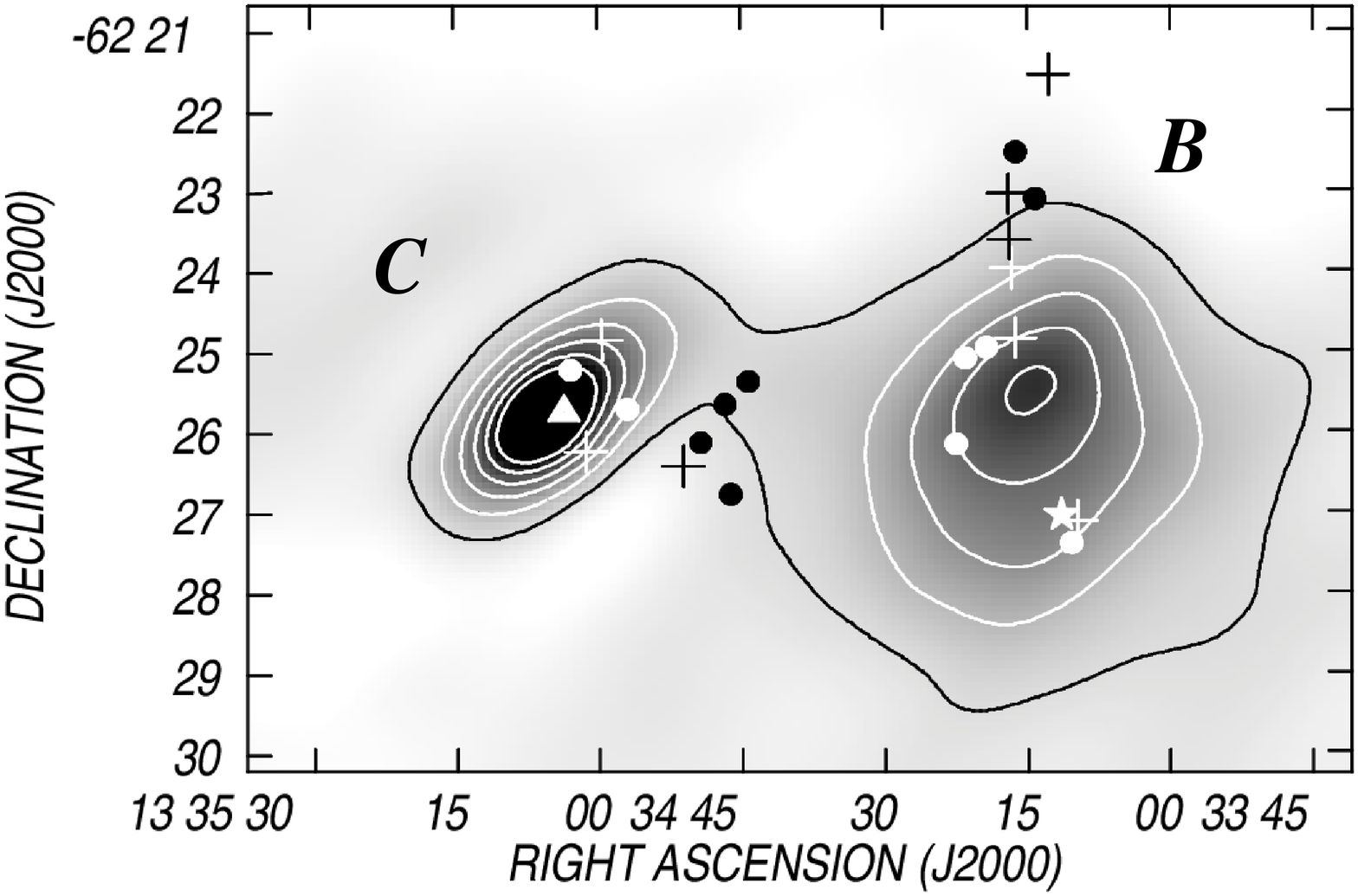}
\caption{{\it Upper panel:} DSS-R image of RCW\,78. The cross marks the position of the WR star. The box encloses the two star forming regions analyzed in this paper. {\it Bottom panel:} The two star forming regions at 60 $\mu$m  (IRAS data). The grayscale goes from 200 to 500 MJy  sr$^{-1}$, and the contour lines are from 240 to 300 MJy sr$^{-1}$ in steps of 20 MJy sr$^{-1}$, and from 350 to 500 MJy  sr$^{-1}$ in steps of 50 MJy sr$^{-1}$. Regions B and C are indicated. The different symbols mark the location of candidate YSOs  identified in Sect. 6: IRAS (star), CHII (triangle), 2MASS sources (crosses), and Spitzer sources (filled circles).}
\label{iras}
\end{figure}

Our targets in this study are two star forming regions probably located in the expanding envelope of the ring nebula RCW\,78 \citep[ from hereon CRMR09]{cappaetal09}. To shed some light on its star-formation capabilities, we have investigated  the presence of dust clumps coincident with star forming regions in the  molecular  layer that surrounds the bubble.

\begin{figure}
   \centering
\includegraphics[width=6.5cm,angle=-90]{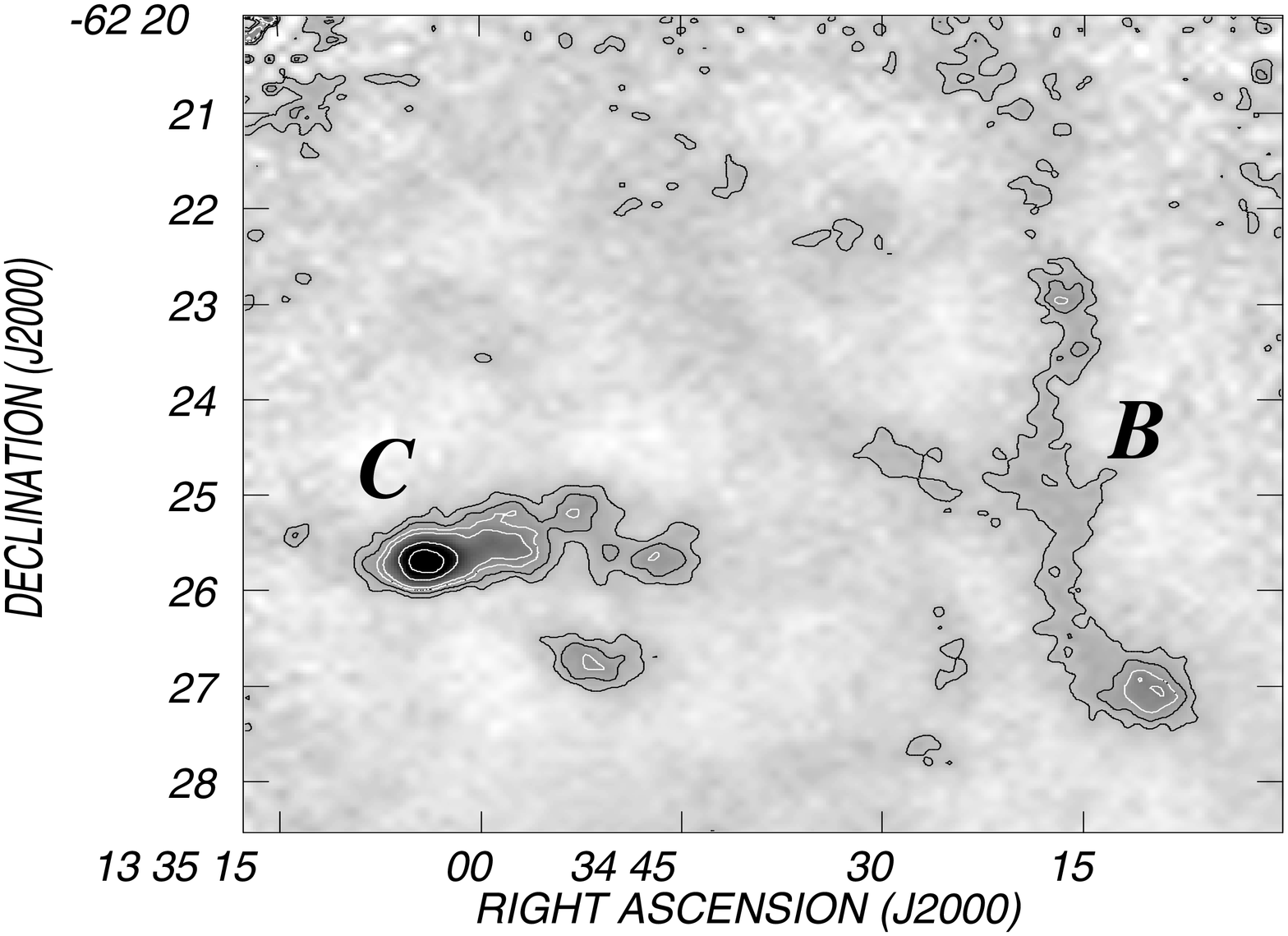}
\caption{870 $\mu$m continuum emission map obtained with LABOCA. The grayscale goes from -10 to 250 \mjyb. Contour levels correspond to 25, 50, 75, 100, 200, and 300 \mjyb.  Regions B and C are indicated.
 } 
 \label{laboca-todo}
 \end{figure}

To unveil  the presence of molecular and cold dust clumps coincident with these star forming regions, we performed \hbox{$^{13}$CO(2-1)} line and sub-millimeter dust continuum observations using the Atacama Pathfinder Experiment telescope (APEX)\footnote{APEX is a collaboration between the Max-Planck-Institut  fur Radioastronomie, the European Southern Observatory, and the Onsala Space Observatory.}, located at Llano de Chajnantor, in the north of Chile. The distribution of cold dust was also investigated using {\em Herschel}\footnote{{\em Herschel} is an ESA space observatory with science instruments provided by European-led Principal Investigator consortia and with important participation from NASA.} images. The dust continuum observations will allow the identification of dust clumps linked to the star forming regions, while molecular line data provide kinematical information  useful to confirm the association of the dense clumps with the neutral layer around the  bubble, as well as to estimate masses and densities.

\section{RCW\,78 and the two star forming regions}

RCW\,78  is a ring nebula of about 35\arcmin\ in diameter related to the Wolf-Rayet star HD\,117688 (= WR\,55 = MR\,49). 
The optically brightest part of RCW\,78 is about 10\arcmin $\times$6\arcmin\ in size and offset to the northwest of the star, while fainter regions are present to the northeast, east, and south \citep[e.g.][CRMR09]{chu+treffers81}. The upper panel of Fig.~\ref{iras} shows the optical image of the nebula. 

\citet{chu+treffers81} found that the velocity of the ionized gas towards the brightest section of RCW\,78 is in the range --53 to --38 \kms. \citet{georgelinetal88} identify H$\alpha$ emission at --41.4 \kms, compatible with Chu \& Treffers's results, and extended diffuse emission at --22 \kms, most probably unrelated to RCW\,78. Circular galactic rotation  models \citep[e.g.][]{brand+blitz93} predict that gas having velocities  between --53 to --38 \kms\  lies at kinematical distances of 3.5-7.0 kpc. Taking into account the $K_s$-value for WR\,55 from the 2MASS catalogue \citep{cutrietal03b}, an absolute magnitude $M_{Ks}$ = --5.92 mag for WN7-9 \citep{crowtheretal06}, and interstellar extinction values from \citet{marshalletal06}, a distance in the range 4.5-5.0 kpc can be derived for the WR star, compatible with the kinematical distance of the ionized gas. Following CRMR09, we adopt a distance  $d$ = 5.0$\pm$1.0 kpc.

CRMR09 presented a study of the molecular  gas associated with the ring nebula RCW\,78 with the aim of analyzing its distribution  and  investigating its energetics. The study was based on  $^{12}$CO(1-0) and $^{12}$CO(2-1) observations of the brightest section of the nebula, carried out with the SEST telescope, and on complementary $^{12}$CO(1-0) data of a larger area obtained with the NANTEN telescope with  an angular resolution of 2\farcm 7.

These authors reported the detection of molecular gas having velocities in the range --56 to --33 \kms\ mainly associated with the western  bright region of RCW\,78,  as well as an \hi\ envelope of the molecular gas, which is described in the same paper. The bulk of the molecular emission appears concentrated in two structures having velocities in the range --52.5 to --43.5 \kms\ and from --43.5 \kms to --39.5 \kms. Gas in the former velocity range is clearly linked to the western section of the nebula. CRMR09 believe that material in the second velocity interval, which partially coincides with the dust lane present at R.A.(J2000) = --62$\degr$ 22$\arcmin$, is also associated with RCW\,78 based on the presence of H$\alpha$ emission probably belonging to the nebula at these velocities (cf. \citealt{chu+treffers81}). This  material may be connected to the receding part of an expanding shell linked to the nebula. Finally, material in the range --39 to --33 \kms\ was only identified in a small region towards the brightest part of RCW\,78, using SEST data. 

Later on, \citet{duronea+12} (from hereon DAT12) performed a study of the molecular gas linked to the western and brightest section of the nebula. They based their analysis on NANTEN data having higher velocity resolution than CRMR09 and found molecular material linked to the western part of the nebula with velocities in the range --54 to --46 \kms\ conforming an expanding ring-like structure whose inner face is being ionized by the WR star. 

A search for candidate young stellar objects (YSOs) performed by CRMR09 using the IRAS, MSX, and Spitzer point source catalogues, resulted in the detection of a number of candidates in two particular areas, suggesting the existence of two  star forming regions (named regions B and C in CRMR09, and showed in the bottom panel of Fig.~\ref{iras}). These two areas coincide with molecular gas belonging to the neutral gas envelope detected around RCW\,78, and then, the possibility that the expansion of the bubble has triggered star formation activity in the dense expanding envelope can not be discarded. 

The star forming regions are centered at \radec\ = (13$^h$34$^m$15$^s$, --62$\degr$ 26$\arcmin$) (named Region B in CRMR09) and  at \radec\ = (13$^h$35$^m$05$^s$, --62\degr 25\arcmin 30\arcsec) (Region C). Both regions are strong emitters at 60 $\mu$m, as can be seen in the bottom panel of Fig.~\ref{iras}.  The O8Ib(f) star  HD\, 117797 (\radec\ = (13$^h$34$^m$11.98$^s$, --62$\degr$ 25$\arcmin$ 1\farcs 8) \citep{walborn82}, located at $d \simeq$ 4 kpc appears projected onto Region B, which also coincides with the open cluster of A- and F-type stars C1331-622 placed at 820 pc \citep{turner+forbes05}. 
 Region C was previously catalogued  as a star forming region by \citet{avedisova02}. 
          
\section{Observations}

\subsection{Continuum dust observations}

\subsubsection{LABOCA image}

To accomplish this project we mapped the sub-millimeter emission at 870 $\mu$m (345 GHz) in a field of 8\arcmin $\times$8\arcmin\ in size centered at  \radec\ = (13$^h$34$^m$30$^s$, --62\degr 26\arcmin) with an angular resolution of 19\farcs 2 (HPBW), using the Large Apex Bolometer Camera (LABOCA) \citep[][]{siringoetal09} at 870 $\mu$m (345 GHz) operating with 295 pixels at the APEX 12-m sub-millimeter telescope.

The field was observed during 1.9 hr in December 2009. The atmospheric opacity was measured every 1 hr with skydips. Atmospheric conditions were very good ($\tau_{zenith} \simeq$ 0.25). Focus was optimized on Mars once during observations. Mars was used as primary calibrator, while the secondary calibrator was IRAS13134--6264. The absolute calibration uncertainty is estimated to be 10\%, 

Data reduction was performed using CRUSH software\footnote{http://www.submm.caltech.edu/~sharc/crush/index.html}. The continuum emission map obtained with LABOCA is shown in Fig.~\ref{laboca-todo}. The  noise level is in the range 10-15 \mjyb. Emission corresponding to Regions B and C is indicated. Signal to noise for  Region B is $S/N \simeq$ = 10, with the higher value corresponding to the  bright southern condensation. For Region C, $S/N$ = 40 at the peak position.

\subsubsection{Herschel images}

The far-infrared ({\sc FIR}) images from {\em Herschel Space Observatory} were also used to trace cold dust emission. We use archival data at  70 and 160\,$\mu$m taken with the Photodetector Array Camera and Spectrometer (PACS; \citealt{poglitsch+10}) and data at 250, 350, and 500\,$\mu$m obtained  with the Spectral and Photometric Imaging REceiver (SPIRE; \citealt{griffin+10}) observed by {\em Herschel} for the Hi-GAL key program (Hi-GAL:{\em Herschel} Infrared GALactic plane survey, \citealt{molinari+10},  OBSIDs: 1342203055 and 1342203086). Both {\em Herschel} imaging cameras were used in parallel mode at 60 arsec/s satellite scanning speed with the purpose to obtain simultaneous 5-band coverege 2\degr $\times$2\degr\ field. The field we use is approximately  centered at [{\em l,b}]\,=\,[--59\degr,0\degr].

\begin{figure*}
\centering
\includegraphics[width=15cm]{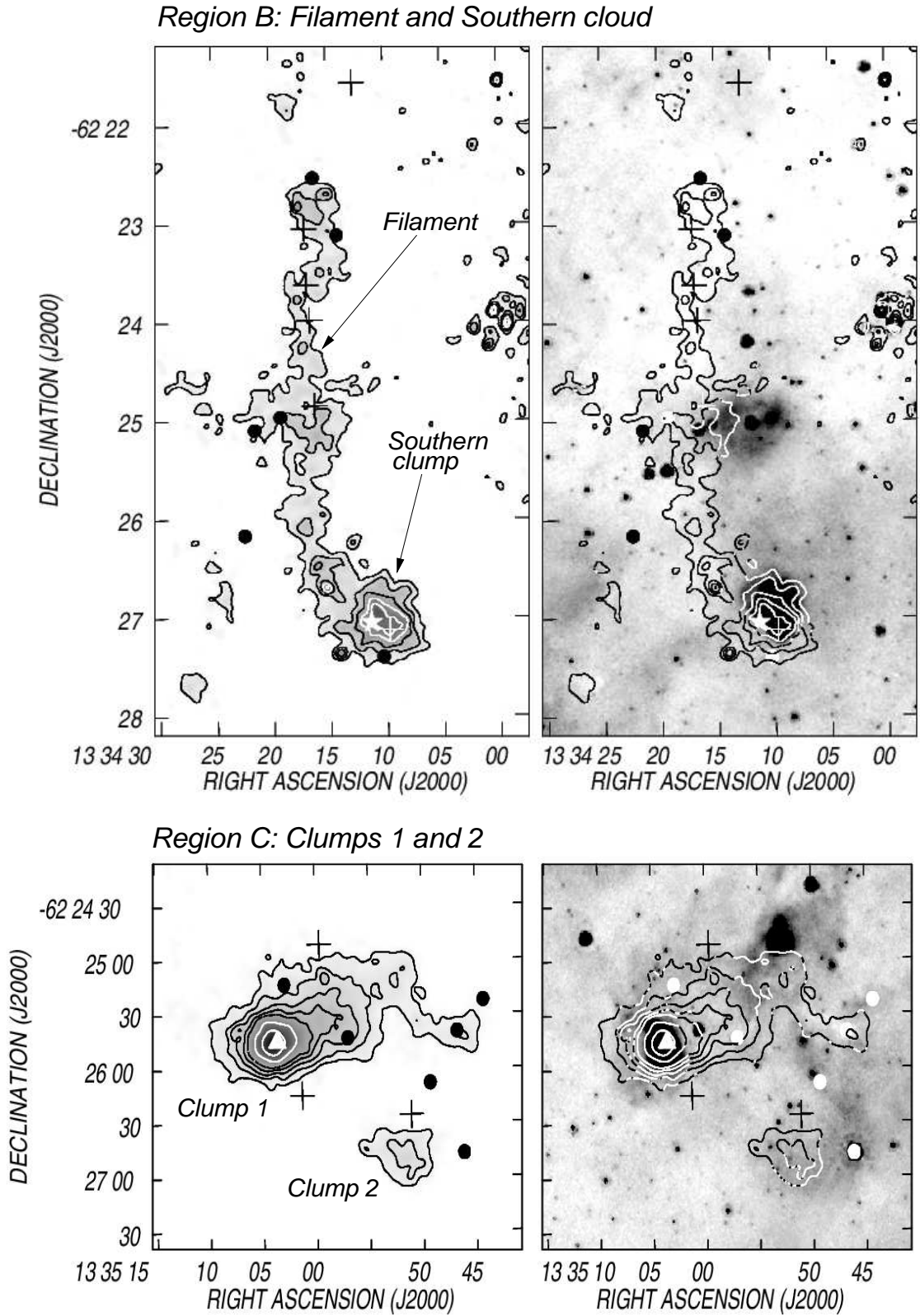}
\caption{Dust emission at 870 $\mu$m corresponding to  Regions B and C. {\it Upper left panel:} Dust emission for the Filament and the Southern clump  (Region B) in contours and grey scale. The image is smoothed to 25\arcsec. The greyscale goes from 10 to 150 \mjyb, and the contours correspond to 20 to 100 \mjyb\ in steps of 20 \mjyb. The position of candidate YSOs is indicated: 2MASS sources (crosses) and Spitzer sources (filled circles) (see Sect. 6). The position of the IRAS source is indicated by a star. {\it Upper right panel:} Overlay of the IRAC emission at 8 $\mu$m in grayscale and the 870 $\mu$m image in contours. The grayscale goes from 28 to 50 MJy ster$^{-1}$.
{\it Bottom left panel:} Dust emission for Clumps 1 and 2  (Region C) in contours and grey scale. The image is smoothed to 25\arcsec. The greyscale goes from 10 to 500 \mjyb, and contours correspond to 40 to 200 \mjyb\ in steps of 40 \mjyb, and 300 and 400 \mjyb. {\it Bottom right panel:} Overlay of the IRAC emission at 8 $\mu$m in grayscale and the 870 $\mu$m image in contours. The grayscale goes from 28 to 50 MJy ster$^{-1}$. The different symbols have the same meaning as in the upper panels. The position of the candidate CHII region is marked by a triangle. 
 }
\label{laboca1}
\end{figure*}

Data reduction from archival data from Level\,2 stage for PACS and SPIRE maps was carried out using the {\em Herschel} interactive processing environment (HIPE v10, \citealt{ott+10}) using the reduction scripts from standard processing. We made a non-oversampled mosaic for PACS images and  used a map merging script to merge two observations, one for each scan direction, performed in SPIRE parallel mode to produce a single map. To reconstruct the three SPIRE maps, we run destriper gain corrections (HIPE v10, updated version by Schulz 2012) using the updated Level\,1 SPIRE data.

To convert intensities from monochromatic values of point sources to monochromatic values of extended sources, assuming that an extended source has a spectral index  alpha = 4  (2 for the blackbody emission plus 2 for the opacity of the dust), the obtained fluxes of each source were multiplied by 0.98755, 0.98741 and 0.96787 for 250, 350 and 500 $\mu$m, respectively.

The angular resolutions at  70, 160, 250, 350 and 500 $\mu$m are 8\farcs 5, 13\farcs 5, 18\arcsec, 25\arcsec, and 36\arcsec, respectively.

\subsection{Molecular observations}

The molecular gas distribution in the area of Region B was investigated by performing $^{13}$CO(2-1) line observations (at 220.398677 GHz) of a region of 3\farcm 2$\times$8\farcm 2 during  December 2009, with the APEX telescope using the APEX-1 receiver, whose system temperature is $T_{sys}$ = 150 K. 

The  half-power beam-width of the telescope is 28\farcs 5. The data were acquired with a FFT spectrometer, consisting of 4096 channels, with a total bandwidth of 1000 \kms\ and a velocity resolution of 0.33 \kms. The  map  was observed in the position switching mode. The off-source position free of  CO emission is located at \radec\ = (13$^h$33$^m$10.3$^s$, \hbox{--62\degr 2\arcmin 41\arcsec)}.

Calibration was performed using  Mars and  X-TrA sources. Pointing was done twice during observations using  X-TrA, o-Ceti and VY-CMa. The intensity calibration has an uncertainty of  10\%.

The integration time per point was 14 sec. The observed line intensities are expressed as main-beam brightness-temperatures $T_{mb}$, by dividing the antenna temperature $T^{\ast}_A$  by the main-beam efficiency $\eta_{mb}$, equal to  0.75 for APEX-1.   

The spectra were reduced using the CLASS software (GILDAS working group)\footnote{http://www.iram.fr/IRAMFR/PDB/class/class.html}. A linear baseline fitting was applied to the data. The typical rms noise temperature  was  0.1 K ($T_{mb}$). AIPS and CLASS software were used to perform  the analysis.

\subsection{Complementary data}

The millimeter and sub-millimeter data were complemented with infrared data retrieved from the Midcourse Space Experiment (MSX) \citep{priceetal01}, 
Spitzer images at 3.6, 4.5, 5.8, and 8.0 $\mu$m from the Galactic Legacy Infrared Mid-Plane Survey Extraordinaire (GLIMPSE) \citep{benjaminetal03}, and images at  24 $\mu$m from the MIPS Inner Galactic Plane Survey (MIPSGAL) \citep{careyetal05}.

Additional images of the Wide-field Infrared Survey Explorer (WISE) \citep{wrightetal10} satellite at 3.4, 4.6, 12.0, and 22.0 $\mu$m with angular resolutions of 6\farcs 1, 6\farcs 4, 6\farcs 5, and 12\farcs 0 in the four bands  were retrieved from IPAC\footnote{http://www.ipac.caltech.edu}. Also, an image at 1.4 GHz from the Southern Galactic Plane Survey (SGPS) published by DAT12 was used. The image has an angular resolution of 1\farcm 7 and an rms sensitivity below 1 \mjyb\ (\citealt{haverkorn+06}).

\begin{figure*}
   \centering
\includegraphics[width=16cm]{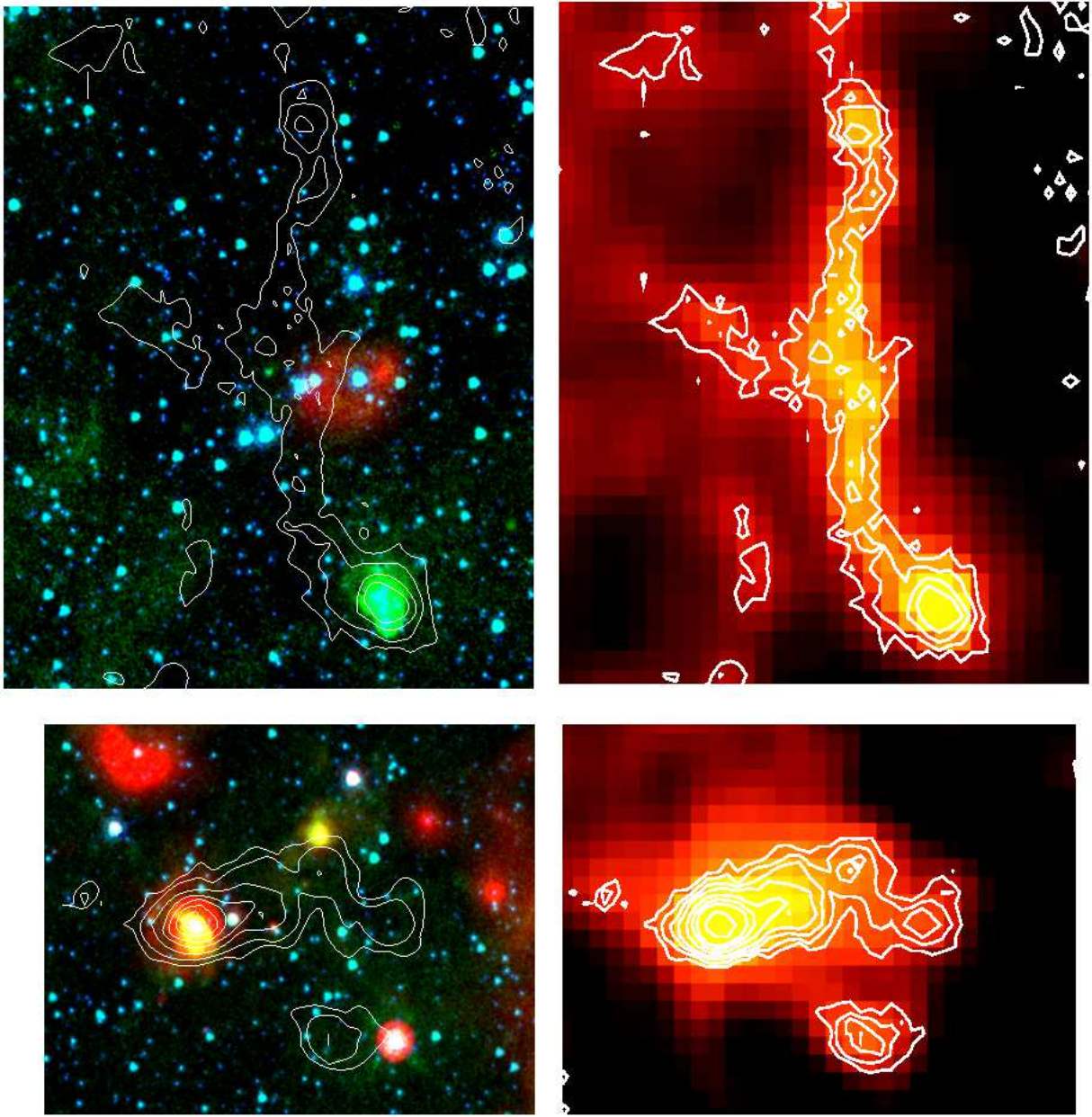}
\caption{{\it Upper left panel.} Composite image showing the IRAC emission of Region B at  4.5 $\mu$m (in blue), 5.8 $\mu$m (in green), 24 $\mu$m (in red), and  870 $\mu$m (white contours). Contour levels correspond to 20, 40, 60, 80, and 120 \mjyb.  {\it Upper right panel.} Composite image showing the Herschel emission of Region B at 160 $\mu$m (in green) and 350 $\mu$m (in red), and the same contour levels of the image on the left.
{\it Bottom left panel.} Composite image showing the IRAC emission of Region C at 4.5 $\mu$m (in blue), 5.8 $\mu$m (in green), 24 $\mu$m (in red), and  870 $\mu$m (in white contours). Contour levels correspond to 20, 40, 80, 120, 160, 200, 300, and 400 \mjyb.  {\it Bottom right panel.} Composite image showing the Herschel emission of Region C at  160 $\mu$m (in green) and 350 $\mu$m (in blue), and the same contour levels of the image on the left.
 } 
 \label{irac-herschel}
 \end{figure*}

\begin{figure}
\centering
\includegraphics[width=8cm]{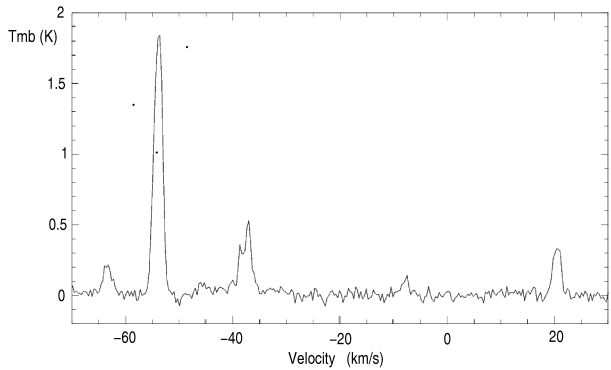}
\caption{$^{13}$CO profile  obtained by averaging the observed spectra in a region of 2\farcm 5$\times$3\farcm 5 centered at \radec\ = (13$^h$34$^m$15$^s$, \hbox{--62\degr 25\arcmin 45\arcsec)}. This region includes the Filament and the Southern clump. Intensity is expressed as main-beam brightness-temperature.
}
 \label{co-prom}
\end{figure}

\section{The distribution of the cold dust and molecular gas}

\subsection{Cold dust distribution}

\subsubsection{Region B}

The upper panels of Fig.~\ref{laboca1} display the emission at 870 $\mu$m in contours and grayscale for Region B (left panel), and an overlay of the cold dust emission  in contours and the IRAC emission at 8 $\mu$m (right panel). The continuum at 870 $\mu$m mainly originates in thermal  emission from cold dust, while the emission at 8 $\mu$m is attributed to polycyclic aromatic hidrocarbons (PAHs) excited by UV photons.

The emission at 870 $\mu$m consists of a filamentary structure elongated along the N-S direction (from hereon the ``Filament''), which ends to the south with a prominent bright condensation (from hereon named ``Southern clump'').  

The Southern clump, centered at \radec\ = (13$^h$34$^m$10.6$^s$, --62\degr 27\arcmin), is 25\arcsec\   in radius (0.60 pc at 5.0 kpc) and  coincides with the brightest area at 8 $\mu$m. In Fig.~\ref{irac-herschel} (left panel), we display the same LABOCA contours of Fig.~\ref{laboca1} superimposed onto a composite image of the IRAC emissions at 4.5 $\mu$m (in red), 5.8 $\mu$m (in green), and  24 $\mu$m (in blue). The emission at 5.8 and 8 $\mu$m are coincident. A number of point-like sources can be identified within this clump at 4.5 $\mu$m, which are described in Sect. 6. Finally, emission linked to the clump is also detected at 24 $\mu$m, coincident with the sources at 4.5 $\mu$m. The right panel of Fig.~\ref{irac-herschel} shows the emissions at 160 (in red) and 350 $\mu$m (in green)  $\mu$m. The image shows the excellent spatial correlation between the Herschel and LABOCA emissions. The clump is brighter at 160 $\mu$m than at 350 $\mu$m. The emissions at 250 and 350 $\mu$m are similar. 

The analysis of the IRAS point source catalogue, which has an angular resolution of 0\farcm 2 to 2\arcmin, allowed the identification of the source IRAS13307-6211 (candidate to YSO/Class 0) as the counterpart at 60 and 100 $\mu$m of the Southern clump. The fluxes and coordinates of this IRAS source are included in Table 1  (see below). The position of the IRAS source is indicated in Fig.~\ref{laboca1} with a  star.

The radio continuum emission distribution at 1.4 GHz shows an extended source (R $\approx$ 1\farcm 7) centered at \radec\ = (13$^h$34$^m$7.5$^s$, --62\degr 26\arcmin 15\arcsec) slightly to the  north of the Southern clump and to the west of the Filament, named CF1 by DAT12. In spite of the low angular resolution of the radio continuum data (100\arcsec) in comparison with the new IR data, the location of the source suggests that it may be related to  the IR bubble described in Sect. 5, as stated by DAT12. However, the extension of the radio source towards the south suggests a contribution from the Southern clump. 

At 870 $\mu$m, the Filament is about 4\farcm 5 in length (6.5 pc at 5 kpc), while its width,  as meassured at 2$\sigma$-level emission varies between 20\arcsec\ and 35\arcsec\ (0.5 and 0.8 pc, respectively), being wider in the middle region. Several relatively  faint maxima can be identified  by eye in the Filament. They are centered at  \radec\ = (13$^h$34$^m$16$^s$, --62\degr 26\arcmin 40\arcsec) (at 6$\sigma$-level emission) and \radec\ = (13$^h$34$^m$16$^s$, --62\degr 25\arcmin 15\arcsec)  (at 4$\sigma$-level emission). These fainter emission regions do not have a clear counterpart at 8 $\mu$m. Indeed, no emission is present either at 4.5 and  5.8 $\mu$m or at 24  $\mu$m (Fig.~\ref{irac-herschel}), suggesting that  excitation sources lack in the filament. The strong emission region at 24 $\mu$m centered at \radec\ = (13$^h$34$^m$12$^s$, --62\degr 25\arcmin) is analyzed in Sect. 5 in connection to the small IR bubble. 
 
As regards the Herschel emission, the right panel of Fig.~\ref{irac-herschel} shows that the emission at 160 $\mu$m has a remarkable resemblance with the southern section of this Filament (Dec.(J2000) $<$  --62\degr 24\arcmin), while emission at 350  $\mu$m is detected all over the Filament. At this last wavelength, the Filament extends further to the north than our LABOCA observations. There is a very good correlation with the LABOCA emission, although the Herschel emission is more extended in size. 

At Dec.(J2000) $>$ --62\degr 22\arcmin 20\arcsec, the emission at 8~$\mu$m is particularly low. In fact, Peretto \& Fuller (2009) identified an infrared dark cloud (IRDC) at \radec\ = (13$^h$34$^m$16.04$^s$, --62\degr 22\arcmin 52.2\arcsec), less than 10\arcsec\ in size, coincident with a relatively bright section of the Filament at 870  $\mu$m.

\begin{figure*}[!ht]
\centering
\includegraphics[width=16cm]{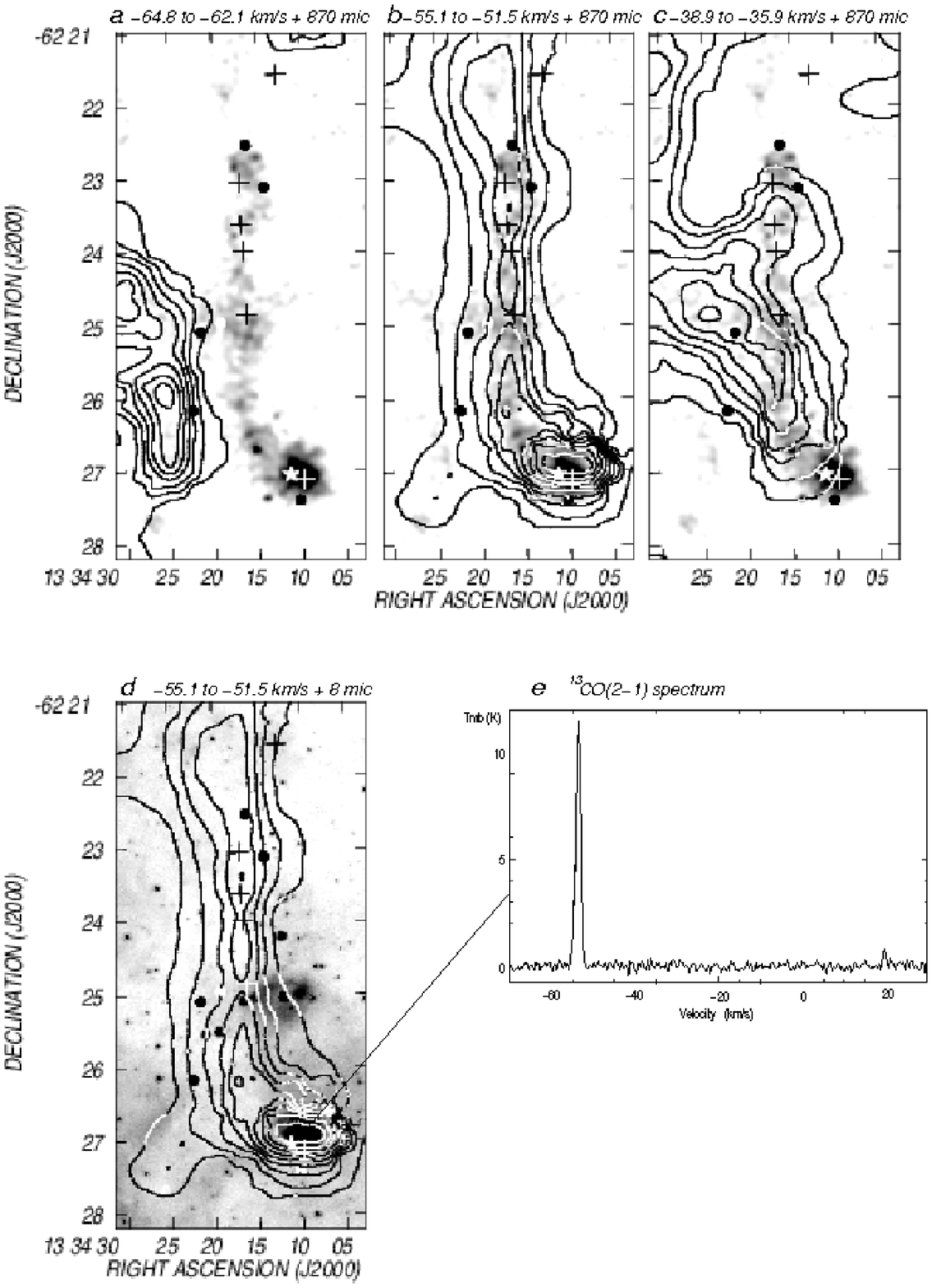}
\caption{Overlay of the integrated emission of the $^{13}$CO(2-1) line within selected velocity intervals (indicated in the upper part of each panel) in contours and the LABOCA image at 870 $\mu$m in greyscale. {\it Panel a.:} Contour lines are from 0.5 to 3.5 K \kms\ in steps of 0.5 K \kms. {\it Panel b.:} Contours are from 1.5 to 12.0 K \kms\ in steps of 1.5 K \kms, and 14.0 K \kms. {\it Panel c.:} Contours are from 0.5 to 3.5 K \kms\ in steps of 0.5 K \kms. 
{\it Panel d.:} Overlay of the $^{13}$CO contours of panel b) and the IRAC 8 $\mu$m emission (in grayscale). 
{\it Panel e.:} $^{13}$CO spectrum towards the brightest region of the Southern clump. 
}
 \label{co-intervalos}
\end{figure*}

\subsubsection{Region C}

The emission at 870 $\mu$m corresponding to Region C is displayed in the bottom panels of Fig.~\ref{laboca1}. The left panel shows this emission in contours and grayscale, while the right panel displays an overlay of the images at 870 and 8 $\mu$m. The image reveals two dust clumps. The brightest one is about 2\farcm 3$\times$1\farcm 0 in size (3.3$\times$1.5 pc at 5 kpc) (from hereon named Clump 1), while the smaller one is 1\farcm 0$\times$ 0\farcm 7 in size (1.5$\times$1.0 pc) (from hereon Clump 2). 

The brightest section of Clump 1 coincides with strong emission at 8 $\mu$m centered at \radec\ = (13$^h$35$^m$4.1$^s$, \hbox{--62\degr 25\arcmin 45\arcsec}) (Fig.~\ref{laboca1}, right panel), with a bright source detected at 24 and 5.8 $\mu$m, and with a point-like source at 4.5 $\mu$m and 3.6 $\mu$m (Fig.~\ref{irac-herschel}, left panel). At 8 and 5.8 $\mu$m, the emission most probably originates in PAHs. Clearly, the emission at 24 $\mu$m suggests the presence of warm dust and an excitation source in the center of Clump 1. The MSX source G307.9563+00.0163, classified as CHII (see Table 1) is projected onto the center of Clump 1 (whose position is marked by a triangle). 

Fig.~\ref{irac-herschel} (right panel) shows an overlay of the Herschel emission at 160 and 350 $\mu$m and  the emission at 870 $\mu$m (in contours). The correlation between  Herschel and LABOCA emissions is excellent. 

Clump 1 coincides with a  radio continuum source detected at 1.4 GHz,  centered at  \radec\ = (13$^h$35$^m$3$^s$, --62\degr 25\arcmin 38\arcsec) indicating the presence of ionized gas. The  presence of radio continuum emission  implies  that a source of UV photons has created a  compact \hii\ region in the inner part of this clump.

A second and fainter source is detected at 8 $\mu$m at \radec\ = (13$^h$35$^m$0.7$^s$, --62\degr 25\arcmin 38\farcs 5), close to the border of the 870 $\mu$m bright region. This source has counterparts at 3.6, 4.5, and 5.8 $\mu$m.  
As regards Clump 2, it is detected at both 160 and 350 $\mu$m. On the contrary, the emission is very faint in the mid IR at 5.8 and 8 $\mu$m. This small region displays weak radio continuum emission at 1.4 GHz. 

\subsection{The molecular gas}

To illustrate the molecular gas components towards the Filament and the Southern clump, we show the  $^{13}$CO(2-1) spectrum averaged in  a region of 2\farcm 5$\times$3\farcm 5  centered at  \radec\ = (13$^h$34$^m$15$^s$, --62\degr 25\arcmin 45\arcsec) in Fig.~\ref{co-prom}. At least four velocity components are detected in the line of sight.  

The bulk of the molecular emission peaks at $\simeq$--53 \kms, while fainter gas components are detected at $\simeq$--63 and --37 \kms. All velocities in this paper are referred to the LSR. The molecular component at --37 \kms displays a double peak structure, while some small scale structure is depicted by the component at --63 \kms. Molecular gas at $\simeq$--8 \kms\ can be barely identified. Finally, molecular emission was detected at  $\simeq$+20 \kms. Molecular gas at --53 and --37 \kms\ was found to be linked to RCW\,78 by CRMR09. As pointed out before, based on morphological arguments, molecular gas at --53 \kms\ is clearly linked to RCW\,78, while the association of the component at --37 \kms might be uncertain according to DAT12.

To investigate the spatial distribution of the molecular gas in the line of sight to the Filament and the Southern clump, we analyzed the $^{13}$CO datacube. Fig.~\ref{co-intervalos} displays an overlay of the integrated emission $I_{CO}$ (= $\int{T_{\rm mb}}\ \ d{\rm v})$ in selected  velocity intervals and the LABOCA 870 $\mu$m emission for comparison. 

The emission in the range --64.8 to --62.1 \kms, which is displayed in panel a), is  unconnected to the continuum emission detected at 870 $\mu$m. 

Panel b) shows gas with velocities in the range --55.1 to \hbox{--51.5} \kms. An overlay of the molecular emission in contours and the dust emission obtained with LABOCA  reveals the excellent correlation between the molecular gas at these velocities and the  dust emission, with the brightest $^{13}$CO emission region coincident with the  Southern clump and  its counterpart detected at 8 $\mu$m (panel d). The CO individual channel maps show that the peak $^{13}$CO emission coincident with the Southern clump is detected in the range --55.1 to --51.5 \kms, while emission having velocities  v $\geq$ --52 \kms\ is detected north of --62\degr 24\arcmin, coincident with the northern section of the Filament. The $^{13}$CO spectrum (panel e) was obtained towards the center of the Southern clump and shows that this velocity component is quite narrow ($\simeq$ 1.3 \kms). The molecular filament, which coincides with the Filament detected at 870 $\mu$m, can be identified in Fig. 5 by CRMR09 (see the panel for the velocity interval between --56.5 to --53.5 \kms) and in Fig. 4 by DAT12, where it is more apparent. 

The molecular gas showed in panel c)  (velocity interval from --38.9 to --35.9 \kms) partially coincides with the Filament and may be also  linked to the nebula.  
However, molecular gas at these velocities is absent in the profile obtained in the line of sight to the southern clump.

To sum up, our analysis confirms the presence of  molecular gas in the  range --55.1 to --51.5 \kms\ clearly associated with Region B, while the association of molecular gas in the interval --38.9 to --35.9 \kms\ remains to be confirmed.

The  molecular gas at v $\simeq$ +20 \kms\ is concentrated towards  \radec\ = (13$^h$34$^m$12$^s$, --62\degr 25\arcmin) and will be discussed in Sect. 5.

As regards Region C, NANTEN data shows that this region coincides with molecular gas in the range --43.5 to --39.5 \kms\ (see CRMR09 and DAT12). The connection of this region to the nebula is an open question  since it is not clear presently that material at these velocities is linked to the nebula.

\begin{figure*}
\centering
\includegraphics[width=6cm]{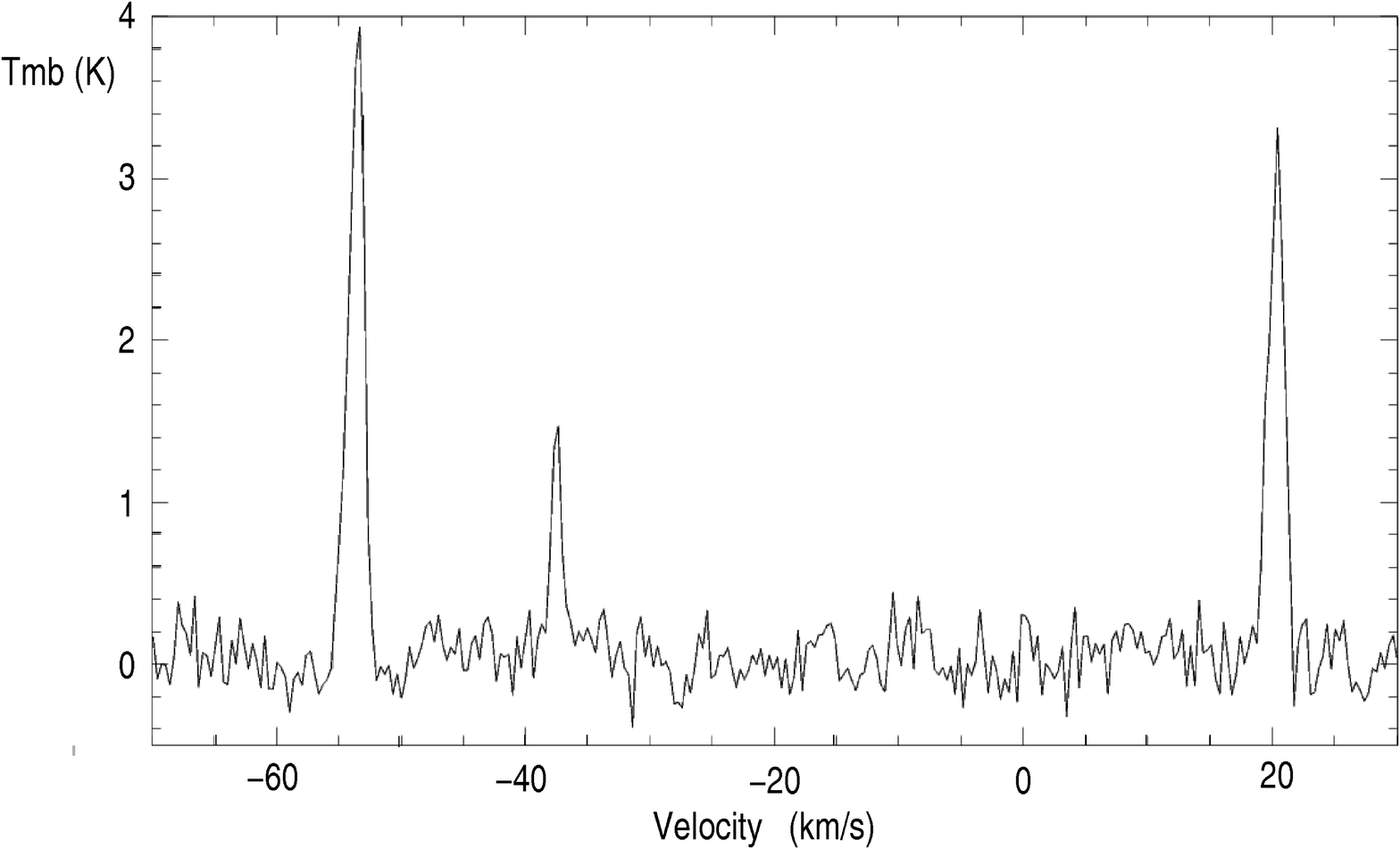}
\includegraphics[width=5cm]{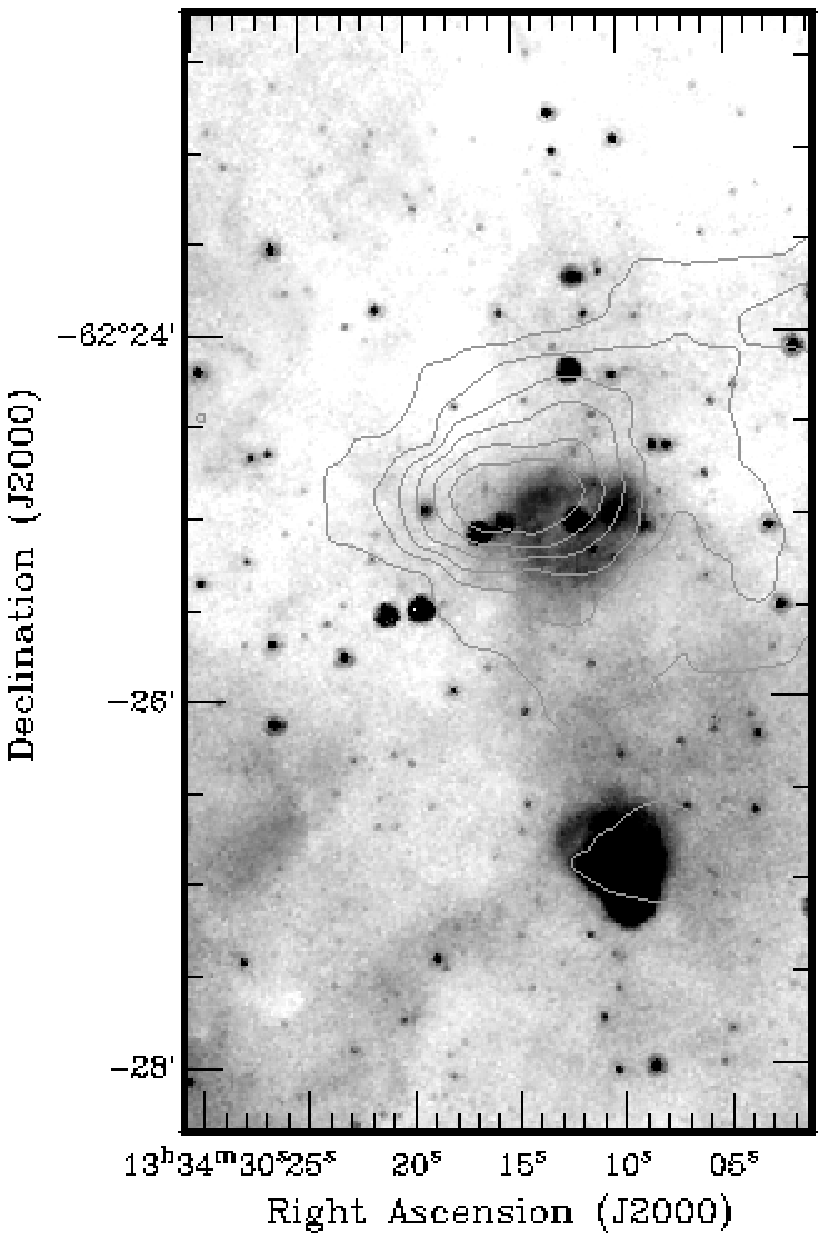}
\includegraphics[width=5cm]{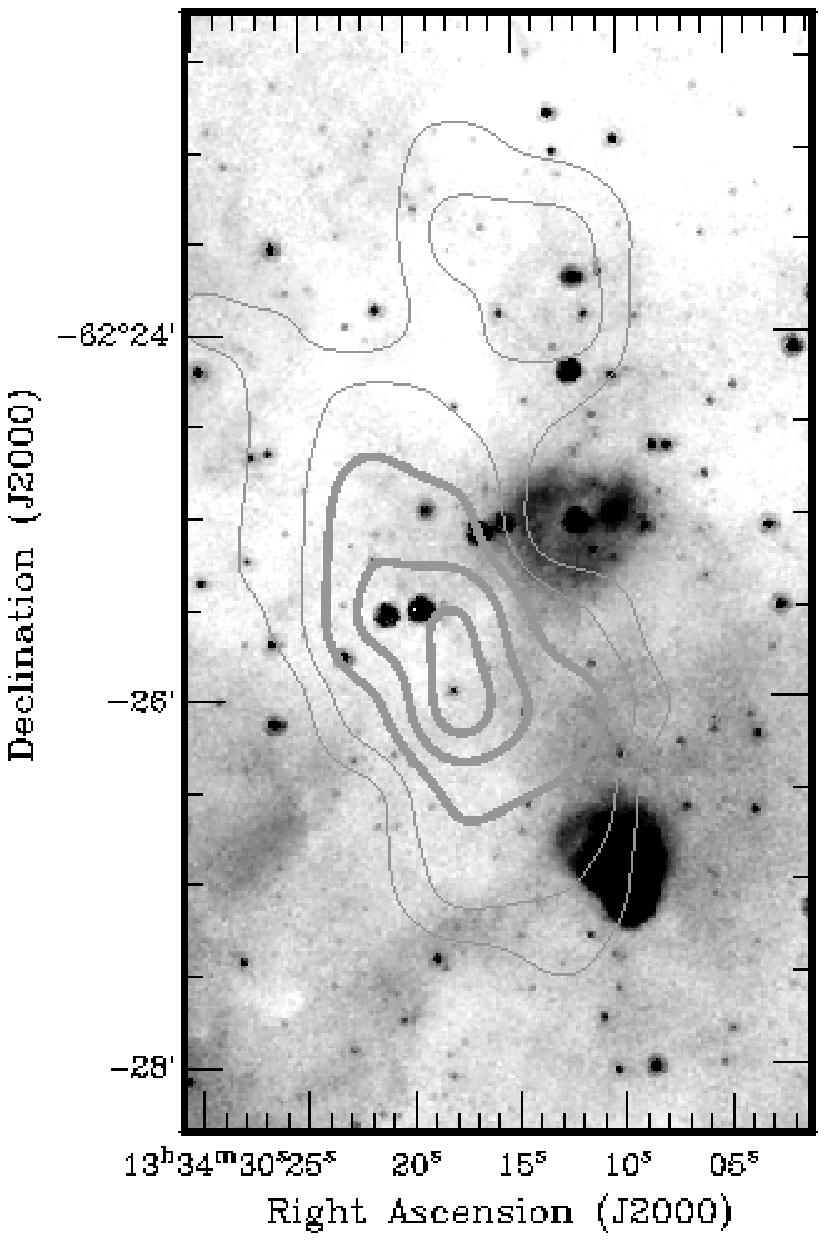}
\caption{{\it Left panel:} $^{13}CO$ spectrum at the center of the molecular gas at +20 \kms\  (\radec\ = (13$^h$34$^m$17.2$^s$, --62\degr 25\arcmin)). Intensity is  expressed as main-beam brightness-temperatures. {\it Middle panel:} Overlay of the integrated emission of the $^{13}$CO(2-1) line within the velocity interval from +19.5 to +21.5 \kms\ and the  emission at 8 $\mu$m from IRAC. Contours are 0.5 to 4.0 K \kms\ in steps of 0.6 K \kms. 
{\it Right panel:} Overlay of the integrated emission in the velocity interval from -37.7 to -37.0 \kms\ and the emission at 8 $\mu$m. Contours are 0.39, 0.46, 1.3, 2.0, and 2.6 K \kms. 
}
 \label{bubble}
\end{figure*}

\section{The small IR bubble}

An inspection of the IRAC image at 8 $\mu$m reveals a striking ring  of about 16\arcsec\ in radius centered at \radec\ = (13$^h$34$^m$12$^s$, --62\degr 25\arcmin). The structure resembles some of the IR dust bubbles identified by Churchwell et al. (2006) at 8 $\mu$m  and \citet{mizuno+10} at 24 $\mu$m. The WISE images at 12 and 22 $\mu$m show an extended source coincident with the IR bubble. A close inspection of the brightest part of this source reveals a faint ring coincident with the bubble, which is also identified at 24 $\mu$m (DAT12)  and in the Herschel image at 70 $\mu$m. The image at 12 $\mu$m includes the emission from prominent PAH features. Emission at  22 $\mu$m (WISE) and 24 $\mu$m (MIPSGAL) indicate the presence of warm dust (\citealt{wrightetal10}). The MSX point source G307.8603+00.0439, listed in Table 1 of CRMR09 as a  candidate CHII region is resolved as the IR bubble in the IRAC-8 $\mu$m image (note that the angular resolution of the IRAC image is 9 times better than that of the MSX image). 
 
In Fig. \ref{bubble} we display a $^{13}$CO spectrum obtained slightly to the east of the IR bubble  at \radec\ = (13$^h$34$^m$17.2$^s$, --62\degr 25\arcmin), showing three intense velocity components. The emission near --53 \kms\ corresponds to the molecular gas linked to the Filament and is unconnected to the IR bubble.  The emission distribution at +20 and --37 \kms\ is displayed in the  central and bottom panels of the figure in contours, with the emission at 8 $\mu$m in grayscale for comparison. 

The bulk of the  molecular emission in the range from +19.5 to +21.5 \kms\ appears projected onto the small IR bubble, while weak emission is also present in the line of sight to the Southern clump. The circular galactic rotation model by \citet{brand+blitz93} predicts that  gas at these velocities should be located outside the solar circle at distances of $\simeq$  11 kpc.
An analysis of the spatial distribution of the \hi\ gas emission at positive velocities using the Southern Galactic Plane Survey \citep{mcclure-griffithsetal05} and that of the molecular gas based on the  $^{12}$CO(1-0) line observations of the NANTEN telescope, reveals the existence of neutral atomic and molecular gas at large scale in the line of sight at v $\simeq$ +20 \kms\ (CRMR09). 

The bottom  panel exhibits the emission of the molecular gas having velocities in interval --37.7 to --37.0 \kms.  The image shows that the $^{13}$CO contours delineate the eastern borders of the IR bubble, suggesting that the nebula is interacting with the molecular material. The existence of molecular emission partially surrounding the IR bubble is compatible with the presence of PAHs emission, which suggests that  a photodissociation region (PDR) developed in the borders of the molecular cloud. 
These facts, along with the existence of warm dust, suggest that an excitation source should be powering the bubble.

HD\,117797 (O8Ib[f], \citealt{maiz-apellanizetal04}), appears projected in the center of the IR bubble. Taking into account an absolute magnitude $M_v$ =  --6.2 mag (\citealt{lang1991}) and an absorption $A_v$ = 2.4 mag derived from photometric data in the GOS catalog (\citealt{maiz-apellanizetal04}), a distance of 4.0 kpc can be estimated. Since HD\,117797 is the only massive star catalogued in this area, as a working hypothesis, we propose that HD\,117797 is the exciting star of the IR bubble, placing it at 4 kpc. We believe that the O8Ib(f) star has created the bubble through the action of its stellar wind and UV photons. 

The IR bubble resembles the  rings identified by \citet{mizuno+10} at 24 $\mu$m, shearing some characteristics with them, in particular its size and its correlation with emission at 8 $\mu$m. According to these authors, a large fraction of these rings would be planetary nebulae. However, the presence of the O8Ib(f) star in the center of the IR structure puts aside a planetary nebula interpretation. 

The MSX source G307.8561+00.0463, at \radec\ = (13$^h$34$^m$10.2$^s$, \hbox{--62\degr 25\arcmin)}, classified as MYSO following  (\citealt{lumsdenetal02}; see CRMR09)  appears projected onto  the ring detected at 8 $\mu$m. This source has a counterpart at 24 $\mu$m. Since this source probably shows  emission from warm dust in the bubble, it is not included in Table 1.

The MSX source G307.8620+00.0399 listed in table 1 from CRMR09 turned out to be a false detection. 

Higher sensitivity and angular resolution molecular and radio continuum data are necessary to  properly investigate if molecular gas is linked to the nebula and to reveal if ionized gas is present. 

\section{Identification of YSOs}

As pointed out in Sect. 1, a search for candidate YSOs in the environs of Regions B and C was performed by CRMR09 using the IRAS, MSX, and Spitzer point source catalogues. The result of the search in the IRAS and MSX catalogues is included in Table 1. 

To search for YSO candidates with better spatial resolution than IRAS and MSX, we used 2MASS sources having good photometric quality ($S/N > 10$), in the (H-K$_S$, J-H)-diagram in  a region of 8\arcmin $\times$8\arcmin\  centered at \radec\ = (13$^h$34$^m$33.75$^s$, \hbox{--62\degr 25\arcmin 34\arcsec)}. We have found 32 candidates YSOs. Nine out of 32 candidate YSOs spatially coincide with regions emitting at 870 $\mu$m. Their coordinates, names and J-, H-, and  K$_S$-magnitudes and MIR-information from Spitzer appear in the middle part of Table 1. Only three sources have been detected in all IRAC bands  (\#5, \#7, and \#11). The upper panel of Fig.~\ref{ysos-allen} shows the 2MASS CC-diagram with the candidate YSOs represented by blue squares. In spite of the small infrared excess of the sources, their location in the diagram is suggestive of being low mass young stellar objects. 

Regarding the Spitzer database, we have performed a new search for candidates YSOs taking into account sources detected in the four IRAC bands following \citet{allenetal04} and \citet{hartmannetal05}, which allows to discriminate IR sources according to a Class scheme: Class I are protostars surrounded by dusty infalling envelopes, Class II are objects whose emission is dominated by accretion disks, Class III are sources with an optically thin or no disk. An analysis of the nature of the Spitzer sources in the same region defined for the 2MASS sources revealed that 29 of 944 IR sources can be classified as candidate YSOs.

The bottom panel of Fig.~\ref{ysos-allen} shows the position of the 29 Spitzer sources in the diagram [3.6]-[4.5] vs. [5.8]-[8.0]. These 29 sources are indicated by grey crosses. Note that because of the revision and reclassification of the  candidates listed in CRMR09, the candidate YSOs identified from the Spitzer point source catalog  listed in Table 1 of this paper differs  from the ones published in CRMR09,  which were mistakenly identified.  

Twelve out of these 29 sources coincide with regions emitting at 870 $\mu$m.  Their IR photometry information is summarized in the bottom part of Table 1. Most of these sources have a counterpart in the 2MASS database. Four out of these 12 sources are represented by bare triangles superimpossed onto the crosses in the bottom panel of Fig.~\ref{ysos-allen}. They correspond to candidate YSOs without a 2MASS counterpart (\#12, \#13, \#17, and \#22), while  filled triangles indicate  sources with only K$_S$ and H and K$_S$ magnitudes  (\#14 and \#20). All but one (\#22) lie in the overlapping region between low-luminosity Class I sources and Class II sources  (\#12, \#13, \#17, and \#20). The other six sources (indicated by asterisks) lie in the limit region between Class II and III sources in the [3.6]-[4.5] vs. [5.8]-[8.0] diagram, which is highly contaminated by evolved stars. Source \#15 has the largest reddenned [5.8]-[8]-color in the sample.

The 2MASS CC diagram of the upper panel of Fig. ~\ref{ysos-allen} indicates that Spitzer sources  \#16,  \#18, \#21, and \#23 (represented by red asterisks) are candidates to giant stars. The other two Spitzer sources with 2MASS counterparts (\#15 and \#19) are situated in the "main sequence star" domain (represented by green filled squares in the upper panel). Note that reddenned main sequence stars and highly obscured YSOs can coincide in the locus of the 2MASS diagram  (\citealt{robitailleetal06}, see Section 3.4.2.1). Spitzer colors corresponding to source \#15 are typical of photodissociation regions (PDR), suggesting its association with molecular gas. 

2MASS candidate YSOs with  Spitzer colors lie in the  region occupied by foreground and  background stars, as well as class Class III stars. They are indicated by bare rectangles in the right panel of Fig.~\ref{ysos-allen}. Sources without measurements at 5.8 and 8 $\mu$m are not included. 

\begin{figure}
\centering
\includegraphics[width=8cm]{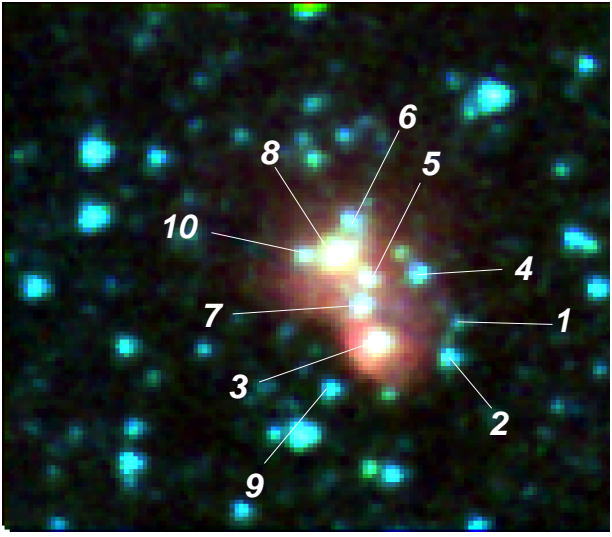}
\caption{ Composite image of the cluster within the Southern clump showing the IRAC emission at 3.6  (in blue), 4.5 (in green) and 5.8 $\mu$m (in red). North is up and east to the left. The size of the image is 82\arcsec\ in R.A. and 75\arcsec\ in Dec. Data for the different sources are included in Table 2. }
\label{cluster}
\end{figure}

Finally, a close inspection of the the IRAC images at 3.6 and 4.5 $\mu$m reveals a cluster of IR sources coincident with the Southern clump. Fig.~\ref{cluster} shows a composite image with the emissions at 3.6 $\mu$m in red, at 4.5 $\mu$m in green, and at 5.8 $\mu$m in blue, with the sources in the cluster. These sources have counterparts in the 2MASS catalogue and  are listed in Table 2, which shows their positions, names, and 2MASS fluxes. The analysis of the 2MASS data allows the identification of one YSO (source \#3 en  Table 1 [=\#3 in Table 2]), one giant star and some main sequence stars. On the other hand, the  Spitzer catalog lists 41 sources projected onto the Southern clump. Only five of them have measurements in the four bands and can not be classified as YSOs. Measured fluxes with HIPE at 3.6 and 4.5 $\mu$m for  the 2MASS sources are indicated in cols. 8 and 9 of Table 2. Sources \#3 and \#8 of Table 2 coincide with point-like emission at 24 $\mu$m, with fluxes of 340 and 48 mJy, respectively. These measurements are compatible with an YSO nature of source \#3. 

\subsection{Characteristics of the YSOs based on their SEDs}

To complete the photometric analysis, we inspected the spectral energy distribution (SED) of the 2MASS and Spitzer sources. We used the online\footnote{http://caravan.astro.wisc.edu/protostars/} tool developed by \citet{robitailleetal07}, which can help to discriminate between evolved stars and reliable candidate YSOs. This tool fits radiation transfer models to observational data according to a $\chi ^2$ minimization algorithm. We choose the models that accomplished the following condition:
\begin{equation}
\chi ^2 \ - \ \chi ^2_{min} \ < \ 3n
\end{equation}
\noindent where $\chi^2_{min}$ is the minimum value of the $\chi^2$ among all models, and $n$ is the number of input data fluxes.

To perform the fittings we used the photometric data listed in Table 1
along with visual extinction values in the range 5-12 mag (derived from the 2MASS CC-diagram), and distances in the range from 4.0 to 6.0 kpc. No observational data longwards than 10 $\mu$m are available for these sources (e.g. MSX, Herschel), except for source  \#7, \#19, and \#22. The available optical  flux for source \#7 was obtained by averaging BVR data from the USNO B-1.0 catalog \citep{zachariasetal04} and the NOMAD catalog \citep{monetetal03}. 

Results of the  analysis applied to the candidate YSOs are summarized in Table 3.
Column 1 gives the number of the source according to Table 1;
 column 2, the value of $\chi^2$ for the best fitting, $\chi^2$/$n$; column 3, the number of input data fluxes used in the fitting, $n$; column 4,  the number of models that satisfies eq. (1), N; column 5, the mass of the central source, $M_{\bigstar}$; column 6, the mass of the disk, $M_{\rm disk}$; column 7, the envelope and ambient density mass, $M_{\rm env}$; column 8, the infall rate, $\dot M_{\rm env}$; column 9 gives the total luminosity, $L_{TOT}$; columns 10 and 11, the  classification according to \citet{allenetal04} and \citet{robitailleetal07}, respectively; and column 12, the age of the central object.

The information in Column 10 is an  estimate of the evolutionary stage of the objects based on their physical properties. Based on ratios of these parameters, the sources can be separated in three categories, including  from objects with significant infalling envelopes and possibly disks (Stage 0/I) to objects with optically thin disks (Stage III). The stages are defined by these cutouts: sources which have $\dot M_{\rm env}$/M$_{\bigstar}$ $>$ $10^{-6}$ yr$^{-1}$ indicate Stage 0/I, $\dot M_{\rm env}$/M$_{\bigstar}$ $<$ $10^{-6}$ yr$^{-1}$ and $M_{\rm disk}$/$M_{\bigstar}$ $>$ $10^{-6}$ correspond to Stage II, and, finally, $\dot M_{\rm env}$/M$_{\bigstar}$ $<$ $10^{-6}$ yr$^{-1}$ and $M_{\rm disk}$/$M_{\bigstar}$ $<$ $10^{-6}$ for Stage III.

Fittings of sources \#14 and \#16 are consistent with Stage I. They have $\dot M_{\rm env} > 0$ and the largest mass envelopes, indicating that they could be young objects inmersed in prominent envelopes. 

Candidate main sequence stars are fitted by Stage III models. For the case of source  \#15, the models suggest that the object is an intermediate mass star (3--5 $M_{\sun}$). However, its position in the 2MASS CC-diagram casts doubts on this interpretation. The SED of  \#19 would correspond to a massive star (8 $M_{\sun}$) with an optically thick disk and an infalling envelope with undetected  accretion.

The results of the fittings should be taken with caution, since the fitting tool has some limitations. As demostrated by \citet{deharveng12} and \citet{offner12} (see also \citealt{robitaille08}), the inferred disk and envelope properties as well as the evolutionary status vary with the viewing angle, the gas morphology, the dust characteristics, and the multiplicity of the sources.

\begin{figure}
\centering
\includegraphics[width=8cm]{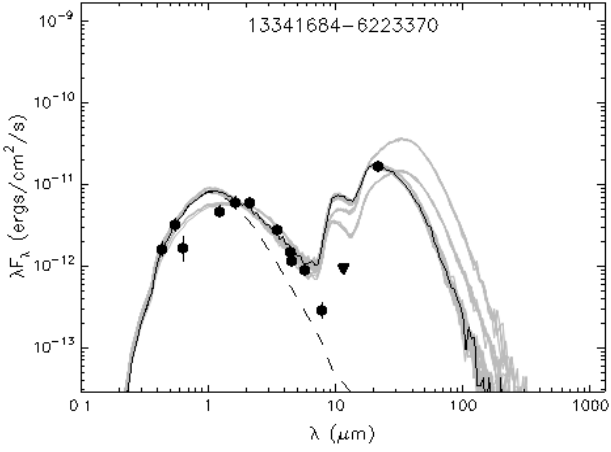}
\caption{SED of the  2MASS source \#7. Input data are indicated by filled circles. The triangle indicates an upper limit at 22 $\mu$m (from WISE). The black line corresponds to the best fitting. The fittings which obey Eq.~1 appear indicated by gray lines. The dashed lines show the emission of the stellar photosphere including foreground interstellar extinction. Fittings for source \#7 indicates a Stage III object, being the oldest source in the sample.}
\label{sed-yso}
\end{figure}

Finally, the fitting of a Kurucz photospheric model gives an additional way to evaluate the nature of sources with NIR colors corresponding to giants in the  (H-K$_S$, J-H) diagram, which lie in the limit of the Class II-III region. The SED of the sources  \#18, \#21, and \#23 can be fitted by a photospheric model with an effective temperature of $\approx$ 5000 K, indicating that they are evolved stars. Consequently, they can be discarded as candidate YSOs and are not included in  Table 3.

The whole sample show colors [3.6]-[4.5] $<$ 1.1, in agreement with moderate or low values of  $\dot M_{\rm env}$. This behavior is related to the accretion rate since the redder the source in [3.6]-[4.5] color, the higher the $\dot M_{\rm env}$ \citep{allenetal04}. 

We applied the SED analysis to the 2MASS candidate YSOs. We were able to discern the evolutionary stage of source  \#7 only, for which we used fluxes from USNO, NOMAD, 2MASS, Spitzer and WISE databases. The fitting suggests that this would be the oldest source with an age of 7$\times$10$^{6}$ yr. The derived SED is shown in Fig.~\ref{sed-yso}.

To summ up, we have not found massive objects in our sample. Source \#16 is consistent with Stages I/II,  while \#7 and \#15 seem to be objects in Stage III with ages larger than 10$^{6}$ yrs. 

\subsection{Spatial distribution}

The position of the YSOs listed in Table 1
is indicated in the bottom panel of Fig.~\ref{iras} and in Fig. \ref{laboca1}. The  star and triangle mark the location of the IRAS source and the compact \hii\ region (CHII), respectively. The IRAS and MSX sources can not be considered as point-like sources.  Circles and crosses correspond to  Spitzer and 2MASS point sources, respectively. 

The IR cluster, which includes source \#3 from Table 1,  coincide with the Southern clump, which is clearly a star forming region. 

A number of sources appear projected onto the filament. Among these sources, source  \#15  may show the presence of molecular gas connected to it. The spatial coincidence of this source with both gas emission at --37 and +20 \kms\ precludes from identifying the associated gas, if any.  Finally, Spitzer sources \#13 and \#14 and the 2MASS sources \#6, \#7 and, \#8 are projected onto  the northern extreme of the Filament. 

Source  \#8 is placed near  the infrared dark cloud IRDC\,307.873+0.079 (\citealt{wilcocketal12}). This IRDC  has  a major axis of 9\arcsec\ in size and  was detected at 250 $\mu$m. \citet{wilcocketal12} suggested that it is an \hii\ region starting to form,  however, no counterparts at wavelengths larger than 4.5 $\mu$m are detected.

As regards Clump 1, the presence of a candidate CHII region and emission at 24 $\mu$m suggests that there was star formation activity.  

\begin{table*}
\centering
 {\scriptsize
\caption[]{Candidate YSOs projected onto Regions B and C. }
\begin{tabular}{cccccccccccc}
\hline
\hline
\multicolumn{3}{l}{\bf IRAS candidate}\\
$\#$ &    [$h \ \ m \ \  s$]    &  [$\circ\ \arcmin \ \arcsec$]    & {\it IRAS} 
&     \multicolumn{4}{c}{Fluxes[Jy]}  & &  & & Classification\\
 & & &name &  12 $\mu$m  &    25 $\mu$m &   60 $\mu$m &  100 $\mu$m & & & &  \\
\hline
1 &  13 34 11.4 & --62 27 03 & 13307-6211 & 0.8  &  2.0  & 20.5 & 85.4   &  &  && YSO/Class 0\\
\hline
\hline
\multicolumn{3}{l}{\bf MSX candidate}\\
$\#$ &   [$h m s$] &  [$\circ\ \arcmin \ \arcsec$] &   {\it MSX} 
&     \multicolumn{4}{c}{Fluxes[Jy]}& & &&  \\     
 &  &  & name &  8 $\mu$m &  12 $\mu$m & 14 $\mu$m & 21 $\mu$m &  & &&Classification \\
\hline
 2 & 13 35 03.6 & --62 25 45 & G307.9563+00.0163 &  0.7808 &  0.7287 & 0.5542 &  3.093  & & & &C\hii\ \\
\hline
\hline
\multicolumn{3}{l}{\bf 2MASS candidates:} \\
$\#$ &  $\alpha$[$h \ \ m \ \ s$]   & $\delta$[$^\circ \ \arcmin\ \arcsec$]  &    {\it 2MASS} &{\it Spitzer }  &\multicolumn{3}{c}{Fluxes[mag]} &\multicolumn{4}{c}{Fluxes[mag]} \\
 &    &    & {\it name}& {\it name}&$J$&$H$&$K_S$& $[3.6]$ & $[4.5]$ & $[5.8]$ & $[8.0]$ \\
\hline
3&13 34 09.79&-62 27 07.2&13340978-6227072&--&14.095&13.364&12.872&--&--&--&--\\
4&13 34 12.71& --62 21 33.9&13341270-6221339& G307.8702+00.1015 &15.059 &14.486 &14.006& 13.726&13.746&--&--\\
5&13 34 16.24&-62 24 50.7&13341623-6224506&G307.8679+00.0464&12.955&12.594&12.232&12.034&11.607&11.518&11.776\\
6&13 34 16.61&-62 23 58.6&13341660-6223585&G307.8708+00.0605&14.400&13.773&13.291&12.602&12.620&12.076&--\\
7&13 34 16.85&-62 23 37.1&13341684-6223370&G307.8724+00.0665&14.788&13.712&12.976&12.313&12.254&12.061&12.331\\
8&13 34 17.01&-62 23 02.9&13341700-6223028&G307.8744+00.0757&15.716&14.881&14.341&13.99&13.75&--&--\\
9&13 34 51.07&-62 26 25.8&13345107-6226258&G307.9297+00.0093&14.949&14.263&13.771&13.703&13.99&--&--\\
10&13 34 59.61&-62 24 51.3&13345960-6224513&G307.9504+00.0323&14.561&14.14&13.809&13.964&13.55&--&--\\
11&13 35 01.30&-62 26 14.7&13350130-6226147&G307.9498+00.0090&12.844&12.415&12.035&11.43&11.479&11.594&11.698\\
\hline
\hline
\multicolumn{3}{l}{\bf Spitzer candidates:} \\
$\#$ &  $\alpha$[$h \ \ m \ \ s$]   & $\delta$[$^\circ \ \arcmin\ \arcsec$]  &   {\it Spitzer } &   {\it 2MASS}&\multicolumn{3}{c}{Fluxes[mag]} &\multicolumn{4}{c}{Fluxes[mag]} \\
 &    &    & {\it name}& {\it name}&$J$&$H$&$K_S$& $[3.6]$ & $[4.5]$ & $[5.8]$ & $[8.0]$ \\
\hline
\hline
12 & 13 34 10.35&--62 27 24.1 &G307.8497+00.0063&--&--&--&--&13.472&12.733&12.344&11.320\\
13 & 13 34 14.14&--62 23 06.7 &G307.8687+00.0756&--&--&--&--&14.070&13.376&12.696&12.109\\
14 & 13 34 16.20&--62 22 31.8&G307.8742+00.0845&13341619-6222319&--&14.534&13.304&12.587&12.458&12.301&11.902\\
15 & 13 34 19.25&--62 24 57.8&G307.8733+00.0435&13341932-6224580&14.572&14.081&13.741&12.985&12.739&11.386&9.771\\
16 & 13 34 21.51&--62 25 05.6&G307.8772+00.0407&13342151-6225058&14.693&13.133&12.466&12.008&11.968&11.932&11.288\\
17 & 13 34 22.47&--62 26 09.9&G307.8761+00.0227&--&--&--&--&13.621&13.084&12.564&11.757\\
18 & 13 34 44.26 & --62 25 22.8 &  G307.9197+00.0286  &13344426-6225229 &   12.565  &11.663&  11.311 & 11.172 & 11.151 & 11.020&  10.635\\
19 & 13 34 46.15 &--62 26 47.1 & G307.9194+00.0049 &13344615-6226472&   10.933 &  8.543 &  7.225 &  6.596 &  6.289 &5.765 &  5.346\\
20 & 13 34 46.74&--62 25 39.9&G307.9236+00.0232&13344675-6225401&--&--&14.204&13.166&12.642&11.946&11.10\\
21 & 13 34 49.29 &-62 26 08.3 &G307.9272+00.0146 & 13344931-6226084  & 13.729 & 12.648 & 12.194 & 12.008  &11.828  &11.964 & 11.569\\
22 & 13 34 56.99&--62 25 43.2&G307.9430+00.0189&--&--&--&--&13.002&11.925&10.894&10.473\\
23 & 13 35 02.93 & --62 25 13.6  &G307.9556+00.0251&  13350291-6225137& 14.425&  13.030 & 12.503 & 12.061 & 12.019 & 12.034&  11.643\\
\hline
\end{tabular}
}
\label{ysos-todos22}
\end{table*}

\begin{table*}
\centering
\caption[]{Cluster of IR sources within the Southern clump. }
\begin{tabular}{rccccccccc}
\hline
\hline
\# &  $\alpha$[$h \ \ m \ \ s$]   & $\delta$[$^\circ \ \arcmin\ \arcsec$]  &    {\it 2MASS} &\multicolumn{3}{c}{Fluxes[mJy](a)} &\multicolumn{2}{c}{Fluxes[mJy]} & Comment \\
&    &    & {\it name}&$J$&$H$&$K_S$& $[3.6]$ & $[4.5]$ \\
\hline
1 &13 34 07.96&-62 27 06.3&13340796-6227062 &0.8  &0.9  &(1.4) & 0.3 & 0.2 \\
2 &13 34 08.53&-62 27 10.9&13340853-6227109 &(0.2)&(0.5)&2.4 & 1.7 & 1.2 \\
3 &13 34 09.79&-62 27 07.2&13340978-6227072 &3.7  &4.6  &4.7  & 2.4 & 2.3 & YSO \\
4 &13 34 08.83&-62 26 58.7&13340882-6226586 &1.9  &5.3  &5.9  & 2.6 &1.4  & giant \\
5 &13 34 09.80&-62 26 58.4&13340980-6226583 &(1.1)&(2.8)&1.9 & 0.7 & 0.7  \\
6 &13 34 09.91&-62 26 50.0&13340991-6226499 &(1.6)&2.0  &3.9 & 1.8 & 1.1  \\
7 &13 34 10.01&-62 27 01.9&13341000-6227018 &(0.8)&(2.4)&3.2 & 1.6 & 1.2   \\
8 &13 34 10.28&-62 26 54.0&13341028-6226540 &5.3  &6.7  &6.2  & 2.3  & 2.5 & MS\\
9 &13 34 10.86&-62 27 13.0&13341085-6227130 &3.2  &3.3  &4.3  & 2.3 & 1.4 \\
10&13 34 10.86&-62 26 53.8&13341086-6226537 &4.3  &4.0  &3.0  & 0.6 & 0.5 & MS \\
\hline
\hline
\end{tabular}
\tablefoot{
\tablefoottext{1}{Values between brackets are uncertain. MS: main sequence star. Source \#3 coincides with source \#3 in Table 1. The other sources do not have a counterpart in Table 1.}\\
}
\label{cluster-tabla}
\end{table*}

\begin{table*}
\centering
\caption[]{Main parameters of the SEDs}
{\scriptsize
\begin{tabular}{cccccccccccc}
\hline
\hline
$\#$ &  $\chi^{2}/n$\tablefootmark{a}  &$n^b$\tablefootmark{b} &N\tablefootmark{c}& $M_{\bigstar}$\tablefootmark{d} & $M_{\rm disk}$\tablefootmark{e} & $M_{\rm env}$\tablefootmark{f} & $\dot M_{\rm env}$\tablefootmark{g} & L$_{\rm TOT}$\tablefootmark{h} &\multicolumn{2}{c}{Classification}\tablefootmark{i} &Age\tablefootmark{j}\\
     &       &       &   & [M$_{\sun}$]&  [M$_{\sun}$]          & [M$_{\sun}$] &  [M$_{\sun}$/yr] & [L$_{\sun}$]   & Class & Stage & [10$^{6}$ yr]\\
\hline
\hline 
7& 15.75&12&37&4--5&5.5 $\times$ $10^{-8}$ -- 1.5 $\times$ $10^{-6}$& 1.81 $\times$ $10^{-4}$  --0.005&0&370--580&III&III&7\\
14 & 1.4& 6&23 & 1 -- 7& 4.2$\times$ $10^{-5}$ -- 0.3 & 0.02 -- 43 & 3.5$\times$ 10$^{-6}$ -- 2 $\times$ $10^{-4}$ &18 -- 480 & II -- III & I &0.1\\
15 & 3.8    &7 &   51& 3 -- 5  & 4 $\times$ $10^{-8}$ -- 3 $\times$ $10^{-6}$ & 4 $\times$ $10^{-8}$ -- 0.04 & 0 -- 1 $\times$ $10^{-9}$ & 130 - 665 & II- III & III& 3\\
16 & 0.3& 7&28& 2 -- 8 & 4$\times$10$^{-4}$ -- 0.4 & 0.02 --16 & 4$\times$10$^{-6}$ -- 5$\times$10$^{-4}$  & 29 -- 566 &II & I&0.1\\
19 & 23&7&1&8& 0.015 &  1.3$\times 10^{-8}$ &0 & 2500& II-III & III&1\\
\hline
\hline
\end{tabular}
\tablefoot{
\tablefoottext{a}{$\chi^{2}$ for the best fitting}; 
\tablefoottext{b}{Number of input fluxes used in the fitting}; 
\tablefoottext{c}{Number of models satisfiyng eq.(1)}; 
\tablefoottext{d}{Mass of the central source}; 
\tablefoottext{e}{Mass of the disk};
\tablefoottext{f}{Envelope mass}; 
\tablefoottext{g}{Infall mass rate}; 
\tablefoottext{h}{Total luminosity};
\tablefoottext{i}{$Class$: classification according to \citet{allenetal04}; $Stage$: classification according to \citet{robitailleetal07}}; 
\tablefoottext{j}{Age of the central object.}
}
}
\label{ysos-robi}
\end{table*}

\begin{figure*}
\centering
\includegraphics[width=0.8\textwidth]{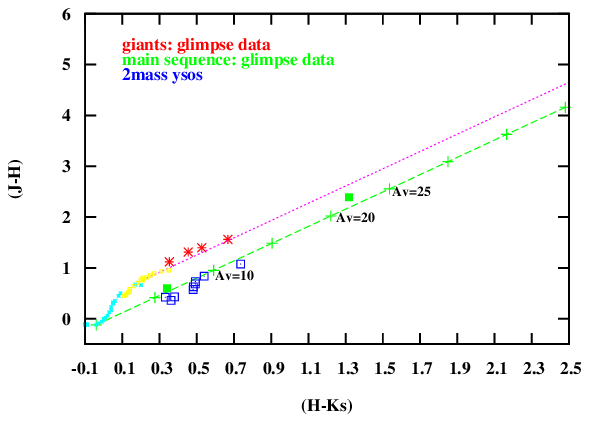}
\includegraphics[width=0.8\textwidth]{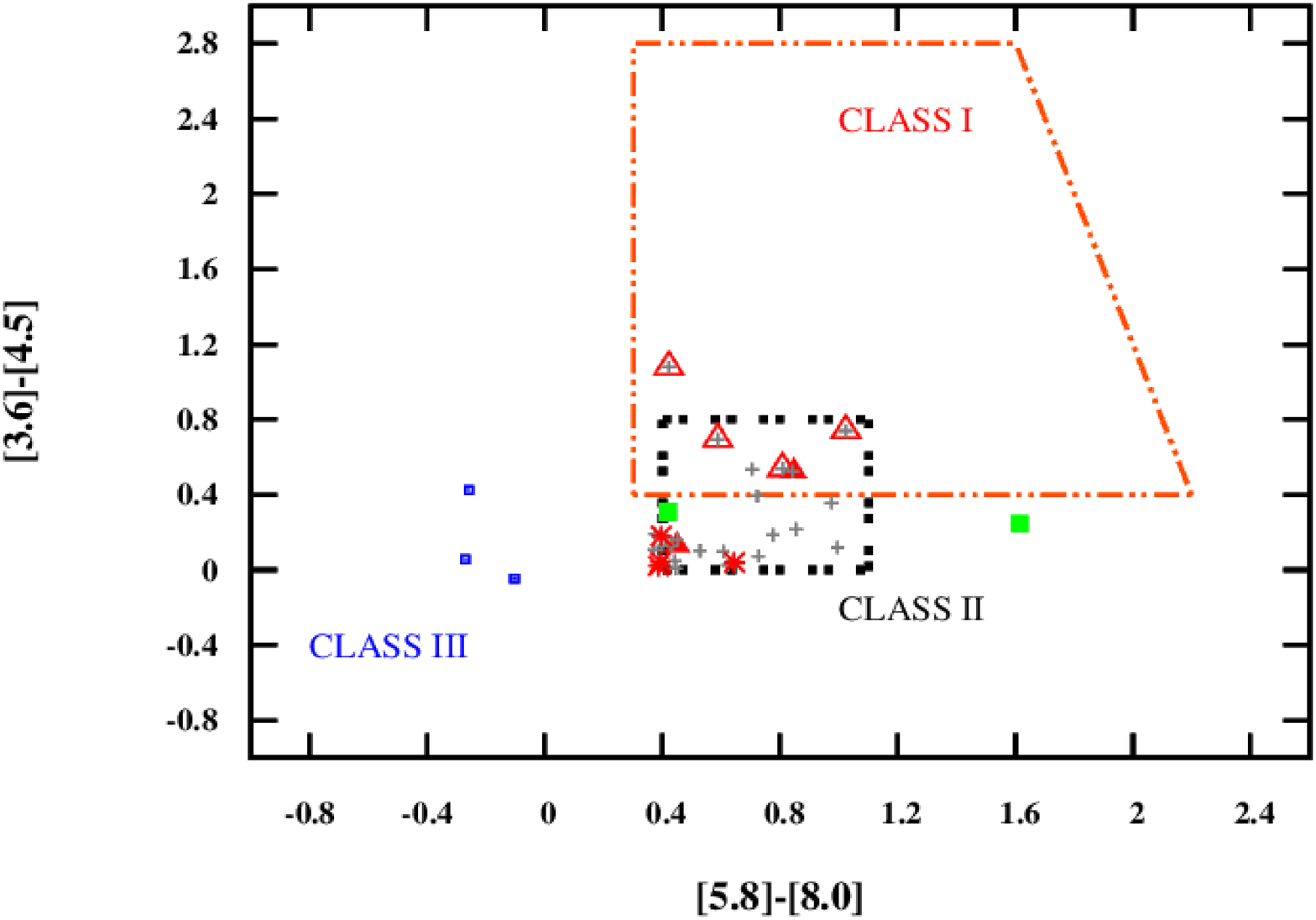}
\caption{{\it Upper panel:} [H-K$_s$, J-H] diagram of the 2MASS and Spitzer candidate YSOs. 2MASS candidate YSOs are indicated by blue squares. 2MASS colors of Spitzer candidates are indicated by asterisks and filled green squares. They are located in the "giant" and "main sequence" regions of diagram.  The dashed lines show the reddening vectors corresponding to  M0 III and  B2 V stars. The crosses are placed at intervals of five magnitudes of visual extinction. 
{\it Bottom panel:} Location of candidates YSOs from the Spitzer database. Crosses represent the 29 sources which lie in Class I and II regions. Triangles, asterisks, and filled rectangles  correspond to sources projected onto the Filament, the Southern clump and Clumps 1 and 2,  respectively. Bare and filled triangles represent candidate YSOs  which do not have a 2MASS counterpart or have been detected in one or two filters only, respectively. Spitzer sources with 2MASS colors are indicated by asterisks and filled rectangles. They occupy the "main sequence" and "giant" regions in the {\it upper panel}. 2MASS sources with Spitzer colors are indicated by bare squares. They lie in the Class III region where main sequence, giants, and pre-main sequence stars overlap. }
\label{ysos-allen}
\end{figure*}

\section{Gas and dust parameters}

\subsection{Masses from submillimeter data }

With the aim of estimating cold dust masses it is necessary to subtract the contribution of different gas phases from the emission at 870 $\mu$m. Two processes may contribute to the emission at this wavelength in addition to the thermal emission from cold dust: molecular emission from the CO(3-2) line and free-free emission from ionized gas. The continuum emission contribution at 345 GHz due to ionized gas was estimated from the radio continuum image at 5 GHz published by CRMR09, considering that the radio emission at 5 GHz is thermal,   using the  expression $S_{345} = (345/5)^{-0.1} S_{5GHz}$. The contribution from the CO(3-2) line was roughly estimated from our CO data taking into account a ratio C0(3-2)/CO(2-1) = 1.0-1.6 \citep{myersetal83} and an intensity ratio $I(^{12}CO/^{13}CO$) = 4-5 for our Galaxy. The total contribution of both mechanisms is about 10\% for the Southern clump and the Filament and less than 5\% for Clumps 1 and 2, and consequently, within the flux uncertainties. 

\begin{figure}
\centering
\includegraphics[width=0.4\textwidth]{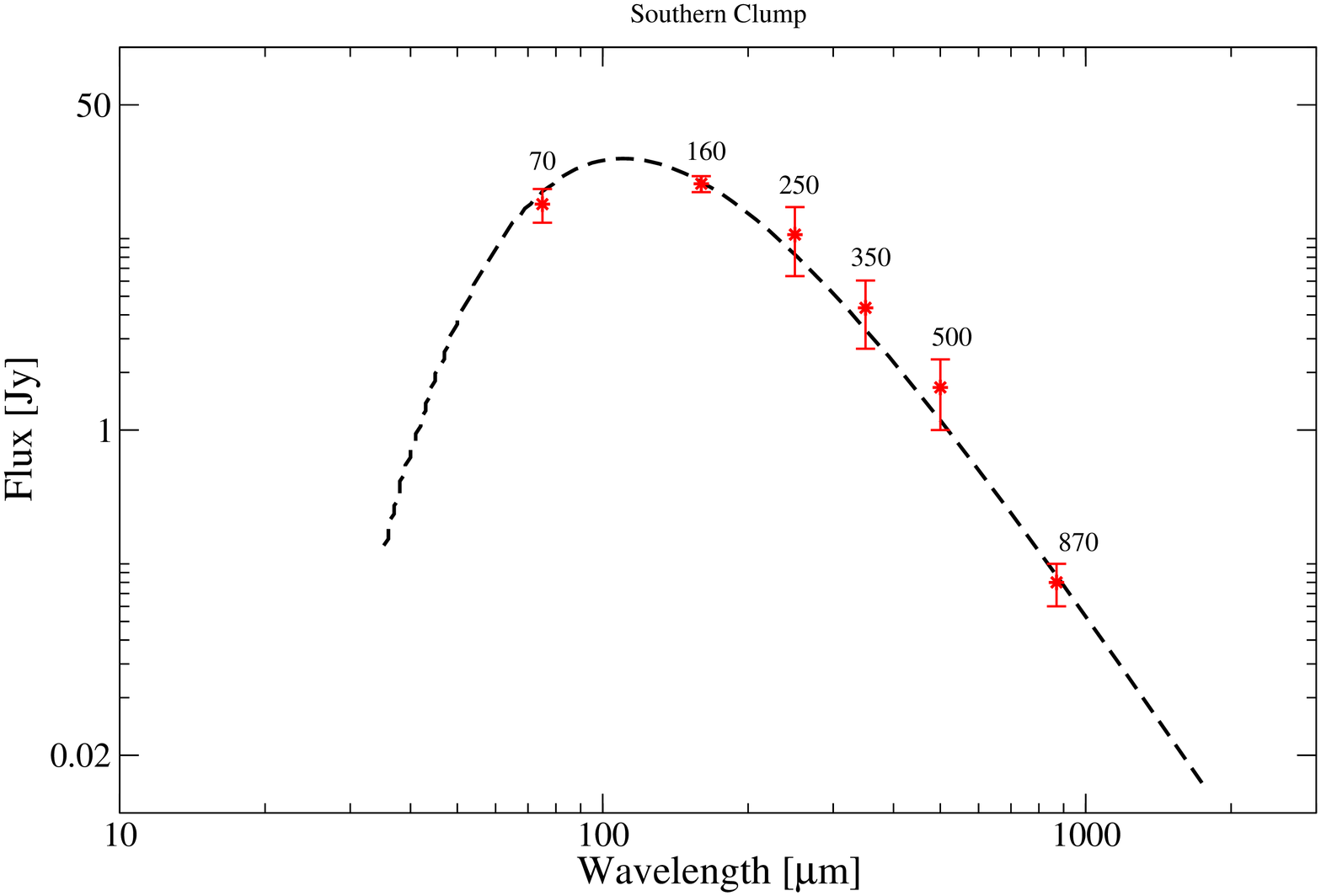}
\includegraphics[width=0.4\textwidth]{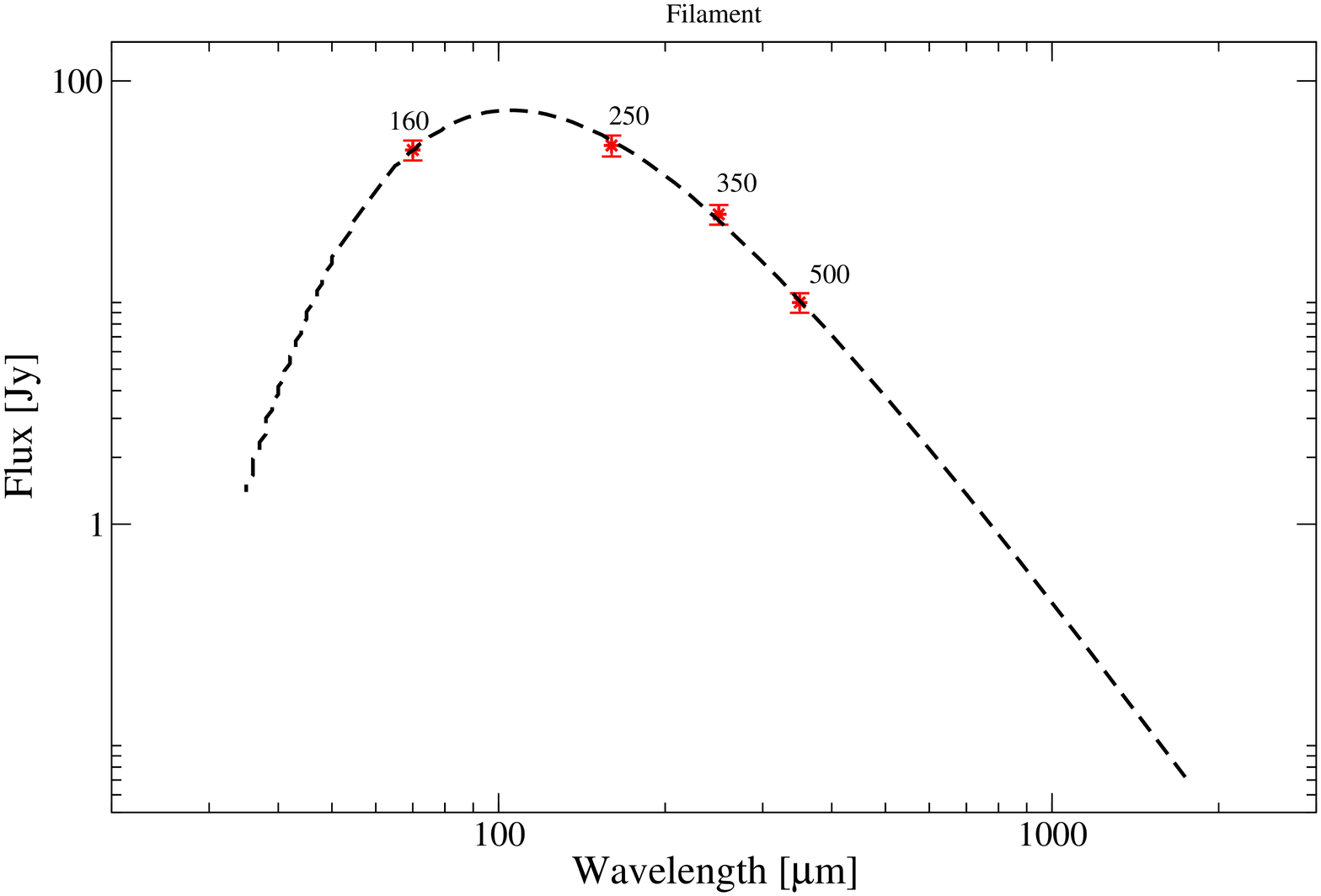}
\includegraphics[width=0.4\textwidth]{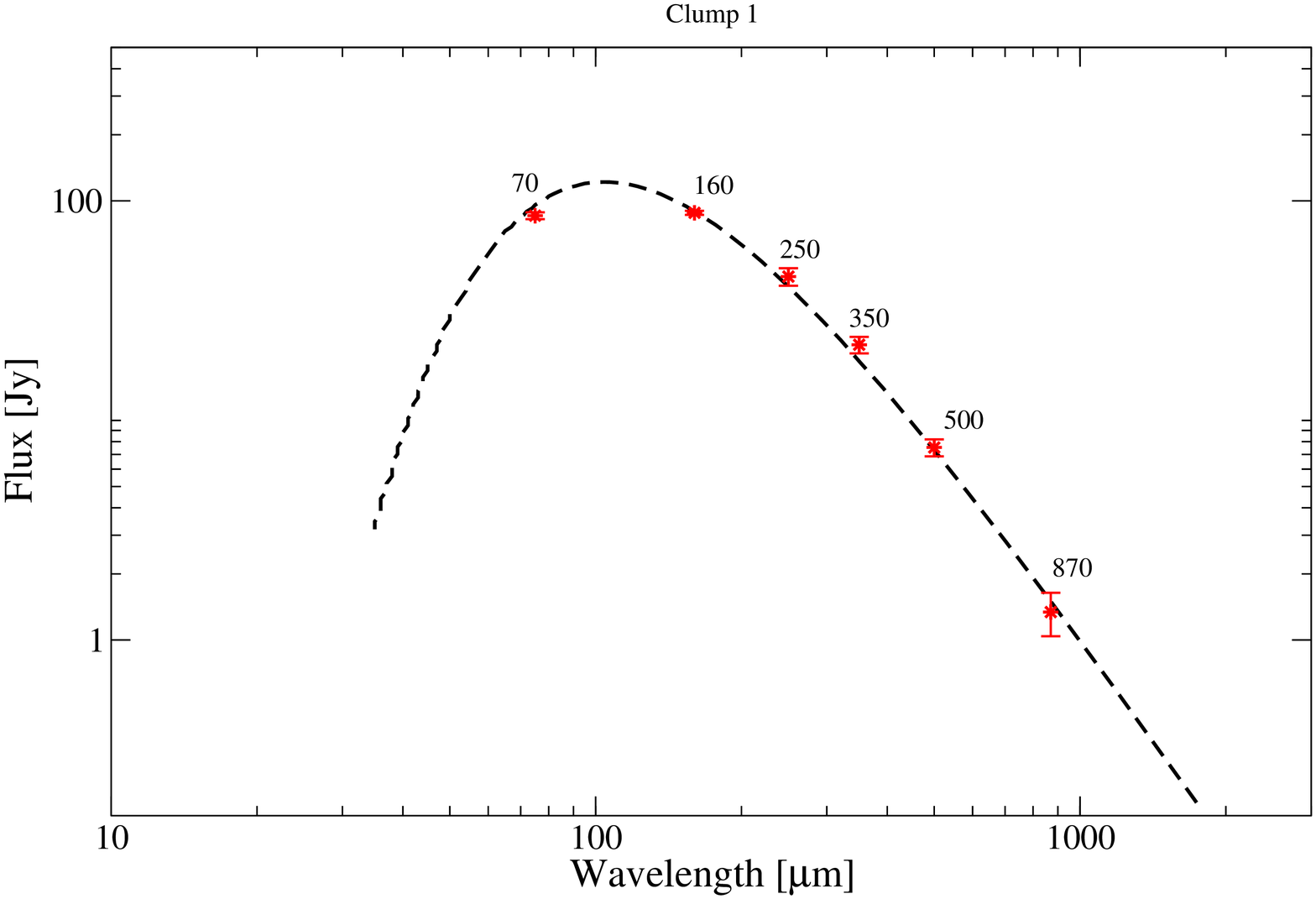}
\caption{SEDs for the Southern Clump (upper panel), the Filament (middle panel), and Clump 1 (bottom panel) obtained using data in the far IR.
}
\label{seds-herschel}
\end{figure}

Considering that the emission detected at 870 $\mu$m originates in thermal dust emission, an estimate of the dust mass can be derived. Assuming that the dust emission is optically thin and following \citet{deharveng+09}, the dust mass can be obtained as:
\begin{equation}
M_{dust} =  \frac{S_{870} d^2}{\kappa_{870} B_{870}(\nu,T_{\rm dust})}
\end{equation}
In this expression, $S_{870}$ is the measured flux density integrated over the area described by the 2$\sigma$ level of the LABOCA map of the source, $d$ is the distance to the source, $\kappa_{870}$ is the dust opacity per unit mass at 870 $\mu$m, and $B_{870}(\nu,T_{\rm dust})$ is the Planck function for a temperature $T_{\rm dust}$. We adopt $\kappa_{870}$ = 1.38 cm$^2$ g$^{-1}$ \citep{miettinen12}.

Dust temperatures $T_{\rm dust}$ were obtained from the Herschel and LABOCA images. Fluxes for the Southern Clump at 70, 160, 250, 350, and 500 $\mu$m were obtained with HIPE using annular apertures sky photometry. To measure the flux, the Filament  was divided into 5 circular sections. For each area, we measured the average surface brightness. Flux uncertainty comes from uncertainties in surface brightness and flux calibration. The results are summarized in Table 4. Columns 2, 3, 4, 5, and 6 list the derived fluxes. Total fluxes are listed in this table for the Filament. 

In Fig.~\ref{seds-herschel} we display the SEDs and the best fitting  obtained for the Southern Clump, the Filament, and the brightest section of Clump 1 after convolving the Herschel and LABOCA images to the angular resolution of the image at 500 $\mu$m and taking into account the same apertures for all the wavelengths. Background emission was substracted from the Herschel images. The  SEDs for the Southern clump and Clump 1 include the flux obtained at 870 $\mu$m.  Derived dust temperature are listed in col. 7 of  Table 4. They are compatible with $T_{\rm dust}$ = 20-30 K, generally  assumed for protostellar condensations \citep[cf.][]{motteetal03,deharveng+09,johnstone+bally06,fontanietal04}. 

\begin{table*}
\centering
\caption[]{Fluxes from Herschel images}
\begin{tabular}{lcccccccc}
\hline
\hline
\\
 &  \multicolumn{5}{c}{Fluxes (Jy)} & $T_d$ \\
 &  70$\mu$m & 160$\mu$m & 250$\mu$m & 350$\mu$m &  500$\mu$m & (K)\\
\hline
\hline 
Southern clump &  15.1$\pm$3.0 & 19.4$\pm$1.9 & 10.5$\pm$4.1 & 4.4$\pm$1.7 & 1.7$\pm$0.7  & 27$\pm$2 \\
Filament       &  --  & 48.8$\pm$5.0 & 51.2$\pm$5.6 & 25.0$\pm$2.6 & 10.0$\pm$1.0 & 31$\pm$3 \\
Clump 1        &  85.6$\pm$3.0 & 88.2$\pm$1.9 & 45.2$\pm$4.1 & 22.1$\pm$1.7 & 7.5$\pm$0.7 & 33$\pm$2 \\
\hline
\end{tabular}
\label{hersch}
\end{table*}

Table~\ref{masses} lists  the parameters derived for the  Southern clump, the Filament, and Clumps 1 and 2, separately, from the molecular and LABOCA data.  In the table we include the size of the source, the flux density, and  dust and total masses.  Fluxes included in this table correspond to the whole extension of each source.  
The dust mass $M_{\rm dust}$ for Clump 2 was obtained using $T_{dust}$ = 20-30 K, while masses for the other structures were estimated using the dust temperature obtained from the SEDs. The total mass $M_{\rm tot}$ was estimated adopting a typical gas-to-dust mass of 100.
 
\begin{table*}
\centering
\caption[]{Dust and molecular gas parameters }
\begin{tabular}{lcccccccc}
\hline
\hline
\multicolumn{3}{l}{\bf LABOCA results:} \\  
& Size  &   $S_{870}$  & $M_{\rm dust}$\tablefootmark{a}  &  $M_{\rm tot}$\tablefootmark{b}  \\ 
  & [arcsec] & [mJy]      & [\msun] &  [\msun] \\ 
\hline
Southern clump &  50           & 1060$\pm$300   & 1.3 &  130  \\
Filament  & 270$\times$27 & 400$\pm$100  & 0.43  &  43 &  \\
Clump 1  & 180$\times$110  &  2520$\pm$340  & 3.2 & 320 & \\
Clump 2 & 30$\times$48    &  230$\pm$50    & 0.24-0.43 & 24--43  \\
\hline
\hline
\multicolumn{3}{l}{\bf Molecular gas parameters:}\\
  & I($^{13}$CO) & $\tau_{13CO}$ & $\Omega$ & R & $N(^{13}CO)$ & $N({\rm H_{2}})$ & $M_{\rm mol}$ &  $n_{\rm H_{2}}$\\
&  [K \kms]&    &  [10$^{-6}$ ster] & [pc] & [10$^{15}$ cm$^{-2}$] & [10$^{20}$ cm$^{-2}$]  & [\msun] &  [ cm$^{-3}$]\\
\hline
Filament     & 3.7  &  0.10 & 0.96 & -- & 2.4$\pm$0.4  &  10.0$\pm$0.4   &  640$\pm$120 & 200 \\
Southern clump   & 5.9   &  0.9 & 0.23 & 1.35 & 5.3$\pm$0.1  & 26$\pm$1   &  340$\pm$10 & 500 \\
\hline
\hline
\end{tabular}
\tablefoot{
\tablefoottext{a}{Derived from the emission at 870 $\mu$m. For the Southern clump, the Filament, and Clump 1 we used $T_{\rm dust}$ derived from the SEDs. For Clump 2, the first value corresponds to $T_{\rm dust}$ = 30 K, and the second one to  $T_{dust}$ = 20 K}\\
\tablefoottext{b}{Derived adopting a gas-to-dust ratio equal to 100.}
}
\label{masses}
\end{table*}

\subsection{Parameters of the molecular gas}

The molecular mass  associated with the Southern clump and the Filament can be evaluated from the  $^{13}$CO line using local thermodynamic equilibrium (LTE). 

Assuming LTE,  the $H_{2}$ column density, $N(\rm {H{_2}})$, can be estimated from the $^{13}$CO(2-1) line data following the equations of \citet{rohlfs+wilson04}
\begin{equation}
N_{tot}(^{13}CO) = 1.5 \times 10^{14}\ \frac{ e^{\left[\frac{T_0
(\nu_{10})}{T_{\rm exc}}\right] }\ \ \ T_{\rm exc} \int \tau^{13}\ \ dv} {1 -
e^{\left[\frac{-T_0 (\nu_{21})}{T_{\rm exc}} \right]} } \qquad \textrm{ (cm$^{-2}$)}
\label{n13co}
\end{equation}
where $\tau^{13}$ is the opacity of the $^{13}$CO(2-1) line, $\nu_{10}$ the frequency of the $^{13}$CO (1-0) line (110.201 GHz), \hbox{$T_0(\nu_{21})$ = $h\nu_{21}/k$}, and $\nu_{21}$ the frequency of the $^{13}$CO(2-1) line. Assuming that \hbox{$\tau^{13} << 1$}, the integral of Eq.
\ref{n13co} can be aproximated by
\begin{equation}
T_{exc} \int{\tau^{13} dv \approx\ \frac{\tau^{13}}{1-e^{(-\tau^{13})}}
\int{T_{\rm mb}}}\ \ dv
\label{integral}
\end{equation}
Then, the molecular mass is calculated using
\begin{equation}\label{eq:masa}
M(\rm H_2)\ =\ (m_{sun})^{-1}\ \mu\ m_H\ \sum\ \Omega\ {\it N}(\rm H_2)\ {\it d}^2 \quad \quad \quad \textrm{(M$_{\odot}$)}
\end{equation}
where m$_{\rm sun}$ is the solar mass ($\sim$ 2 $\times$ 10$^{33}$ g), $\mu$ is the mean molecular weight, which is assumed to be equal to 2.8 after allowance of a relative helium abundance of 25\% by mass \citep{Y99}, m$_{\rm H}$ is the hydrogen atom mass ($\sim$ 1.67 $\times$ 10$^{-24}$ g), $\Omega$ is the solid angle subtended by the CO clump in ster, $d$ is the assumed distance expressed in cm, and $N$(H$_2$) is the H$_2$ column density, obtained using a ``canonical'' abundance \hbox{$N(\rm H_2)$ / $N(^{13} {\rm CO})$} = \hbox{5$\times$10$^{5}$} \citep{dickman78}. Assuming that \hbox{$\tau^{13} << 1$} and \hbox{$\tau^{12} >> 1$} the opacity of the \coc\ line can be derived using
\begin{equation}
\quad \tau^{13} = -{\rm ln}\left[1-\frac{ T_{\rm b}(^{13}CO)}{T_0^{13}}
\left[\left(e^{\frac{T_0^{13}} {T_{\rm exc}}}-1
\right)^{-1}-f(T_{\rm bg})\right]^{-1}\right]
\label{tau13co}
\end{equation}
where $T_0^{13}$ = $h \nu_{13} / k$, $f(T) = \left[ e^{T_0^{13} / T} -1 \right]^{-1}$, and $T_{\rm b} = 2.7$ K. Since the optically thick $^{12}$CO(2-1) line was not observed, T$_{\rm exc}$ could not be derived directly from our data. Hence, we have adopted  T$_{\rm exc}$ = 20-40 K for the Southern clump in agreement with values adopted by \citet{brand+01}, \citet{deharveng+08}, and \citet{pomares+09}. Similar values were adopted for the Filament. It is worth to point out that  uncertainties of 50$\%$ in the values of T$_{\rm exc}$ yields to total $M(\rm H_2)$-uncertainties of up to 25$\%$. 

The results for the Southern clump and the Filament are indicated in Table~\ref{masses}, which lists the integrated emission of $^{13}$CO within the velocity interval from --55.1 to --51.5 \kms, the optical depth, the solid angle and the effective radius where the molecular emission is present, the column density of $^{13}$CO and molecular hydrogen,  the molecular mass, and the ambient density. Errors in masses due to distance uncertainties (20\%) are about 40\%.  The ambient density of the Southern Clump was  estimated by distributing the total mass in a sphere of 1.35 pc in radius,  while for the filament, we distributed the molecular mass in a region of 8 pc in lengh, 2.7 pc in width, and 2.7 pc along the line of sight. 

Note that in the Southern clump, the molecular gas was detected in a larger area  than the cold dust. The difference in the emission areas may be due to the presence of emission below the 2$\sigma$ limit in the LABOCA data and/or a higher gas-to-dust ratio in the outer areas of the Southern clump.
In order to compare the masses derived from the dust continuum $M_{tot}$ and from the molecular line, we need to integrate the emissions within the same area. Consequently, we evaluated the  molecular mass using LTE in the same area where the dust continuum emission was detected (equal to the area of a circle of R = 0.6 pc). We obtained a total mass $M'_{mol}$ = 100 M$_0$. 

We can obtain the virialized mass for this clump and compare it with $M_{tot}$ and $M'_{mol}$. Assuming a spherically symmetric cloud with a constant density distribution, the virial mass  can be determined from \citep{maclaren+88}
\begin{equation}
\label{eq:mvir}
\quad M_{vir}\ =\ 210\ R\ (\Delta {\rm v}_{\rm cld})^2 \qquad \qquad
\text{[\msun]}
\end{equation}
where $R$ = 0.6 pc, 
and $\Delta$v$_{\rm cld}$ = 1.5 \kms\ is its velocity dispersion in \kms. The latter is defined as the FWHM line width of the composite profile derived by using a single Gaussian fitting. The composite profile is obtained by averaging  the spectra within $R$. The virial mass turns out to be $M_{vir}$ = 280 M$_0$. The agreement between the masses derived using the LTE model and  the virial theorem is within a factor of 3, while the agreement of the molecular and virial masses with the mass derived from the dust continuum emission is within a factor of 2-6.  

The mass of the Southern clump is similar to masses derived for other star forming regions (see for example, Mookerjea et al. 2004, \citealt{deharveng+09}, \citealt{sanchez-monge+08}).

\section{Conclusions}

We have investigated the presence of molecular clouds and dust clumps linked to the star forming Regions B and C  previously identified  in CRMR09,  by performing APEX observations of the continuum 870 $\mu$m dust and $^{13}$CO(2-1) molecular line emissions, and analyzing IRAC and Herschel images in the near and far infrared.

The continuum sub-millimeter observations towards Region B allowed the identification of a Filament elongated in the N-S direction, ending in a bright condensation (named Southern clump). This Southern clump,  which can be  identified with IRAS\,13307-6211, is 25\arcsec\ in radius (or 0.6 pc at 5 kpc), and coincides with  bright patches of emission at 5.8 and 8 $\mu$m, and in the Herschel images  at 70, 160, 250, 350, and 500 $\mu$m. The presence of radio continuum emission probably linked to the clump and emission at 24 $\mu$m suggests the existence of excitation sources inside. The identification of a cluster or IR sources detected in the 2MASS and IRAC images, which includes young stellar objects, showed that star formation is active in this region.

The Filament can be identified also in the Herschel images. On the contrary, it is not detected in the near IR. A number of candidate YSOs are projected onto the filament. 

The distribution of the $^{13}$CO emission revealed that molecular gas with velocities in the interval --55.1 to --51.5 \kms\ is the molecular counterpart of the dust Filament and the Southern clump. The velocity of the molecular gas is compatible with its location  in the expanding envelope of RCW\,78. 

Towards Region C, two cold dust clumps were identified both in the LABOCA and Herschel images.  The brightest section of the larger clump coincides with emission at  5.8 and  8 $\mu$m. The  presence of radio continuum emission at 1.4 GHz is indicative of ionized gas and a source of UV photons inside. Indeed, an MSX source classified  as candidate CHII region reinforces the idea that moderate star formation has occured recently. It is not clear presently if this region is linked to RCW\,78. 

We estimated total masses for the dust condensations from the emission at 870 $\mu$m and from the molecular line using LTE and the virial theorem. Masses for the Filament and the Southern clump obtained through different methods agree within a factor of 2-6.  Dust temperatures derived from far IR data for the dust clumps and the Filament are about 30 K, compatible with those of protostellar condensations. 

The 8 $\mu$m-IRAC image revealed the existence of an infrared dust bubble of 16\arcsec\ in radius centered at \radec\ = (13$^h$34$^m$12$^s$, --62\degr 25\arcmin) with a clear counterpart at 24 $\mu$m, suggesting the existence of warm dust in the ring. The bubble is probably linked to the O-type star HD\,117797 located at 4 kpc  and  may be interacting with the molecular gas, although molecular gas linked to it can not be clearly identified from the present observations. The presence of the PAH emission seen at 8 $\mu$m and of warm dust is compatible with the O-type star as the excitation source  powering the bubble.

\begin{acknowledgements}
C.E.C. acknowledges the kind hospitality of Dr. M. Rubio and her family during her stays in Santiago, Chile.  We acknowledge the many constructive comments and suggestions of the referee, which helped to improve this paper. This project was partially financed by CONICET of
Argentina under project PIP 02488 and UNLP under project 11/G120. M.R. is supported by CONICYT of Chile through grant No. 1080335.
V.F. would like to thank Ivan Valtchanov for his support and valuable assistance in Herschel data processing.
This research has made use of the NASA/ IPAC Infrared Science Archive, which is operated by the Jet Propulsion Laboratory, California Institute of Technology, under contract with the National Aeronautics and Space Administration. This work is based [in part] on observations made with the Spitzer Space Telescope, which is operated by the Jet Propulsion Laboratory, California Institute of Technology under a contract with NASA. This publication makes use of data products from the Two Micron All Sky Survey, which is a joint project of the University of Massachusetts and the Infrared Processing and Analysis Center/California Institute of Technology, funded by the National Aeronautics and Space Administration and the National Science Foundation. The MSX mission is sponsored by the Ballistic Missile Defense Organization (BMDO).
\end{acknowledgements}

\bibliographystyle{aa}  
\bibliography{sf-gis6}
{\typeout{}
\typeout{****************************************************}
\typeout{****************************************************}
\typeout{** Please run "bibtex \jobname" to optain}
\typeout{** the bibliography and then re-run LaTeX}
\typeout{** twice to fix the references!}
\typeout{****************************************************}
\typeout{****************************************************}
\typeout{}
}

\end{document}